\documentclass[aps,prb,reprint,showpacs,superscriptaddress,longbibliography]{revtex4-2}
\usepackage{mathtools,amssymb,graphicx,units}

\usepackage[plainpages=false,pdfpagelabels,colorlinks=true,linkcolor=red,urlcolor=blue,citecolor=blue,pdftitle={Title},pdfauthor={},pdfdisplaydoctitle=true,pdfduplex=DuplexFlipLongEdge]{hyperref}
\usepackage{verbatim}
\usepackage[english]{babel}
\usepackage{xcolor}

\newcommand\Mycomb[2][^n]{\prescript{#1\mkern-0.5mu}{}C_{#2}}

\usepackage[normalem]{ulem}

\graphicspath{{./figs/}}

\begin{document}

\title{Deconvolving the components of the sign problem}
\author{S. Tarat}
\email{tarats@csrc.ac.cn}
\affiliation{Beijing Computational Science Research Center, Beijing 100193, China}
\author{Bo Xiao}
\email{bxiao@flatironinstitute.org}
\affiliation{
Center for Computational Quantum Physics,
Flatiron Institute,
New York, New York 10010, USA}
\author{R. Mondaini}
\email{rmondaini@csrc.ac.cn}
\affiliation{Beijing Computational Science Research Center, Beijing 100193, China}
\author{R.T. Scalettar}
\email{scalettar@physics.ucdavis.edu}
\affiliation{Department of Physics, University of California,
Davis, CA 95616, USA}

\begin{abstract}
Auxiliary field  Quantum Monte  Carlo
simulations of interacting fermions require
sampling over a Hubbard-Stratonovich field $h$ introduced to decouple the interactions.
The weight for a given configuration involves the products of the
determinant of matrices
$M_\sigma(h)$, where $\sigma$ labels the species, 
and hence is typically not positive definite.  Indeed, the average sign 
$\langle {\cal S} \rangle$ 
of the determinants goes  to zero exponentially with increasing spatial size
and decreasing temperature for most Hamiltonians 
of interest.  This statement, however, does not explicitly separate two possible
origins for the vanishing of 
$\langle {\cal S} \rangle$.
Does $\langle {\cal S} \rangle \rightarrow 0$
because {\it randomly} chosen field configurations have
${\rm det}\big(M(h)\big) < 0$,
or does the `sign problem' arise 
because the specific subset of configurations chosen by the weighting
function have a greater preponderance of negative values?  In the latter case, the process of weighting the configurations  with
$|{\rm det}\big(M(h)\big)|$ might steer the simulation to a region of
configuration space of $h$ where positive and negative determinants
are equally likely, even though randomly chosen $h$ would preferentially
have determinants with a single dominant sign.
In this paper we address the relative importance of these two 
mechanisms for the vanishing of 
$\langle {\cal S} \rangle$
in quantum simulations.
\end{abstract}

\date{Version 16.0 -- \today}


\maketitle

\section{Introduction}  \label{sec:Introduction}

Auxiliary Field Quantum Monte Carlo (AFQMC) 
\cite{blankenbecler81,white89,chen92,assaad02,gubernatis16,alhassid17,hao19,he19},
relies on the observation that traces over products of exponentials of quadratic forms of fermionic operators can be done analytically. Thus in QMC for a Hamiltonian like the Holstein model, where only such quadratic forms are present, the resulting simulation samples over the remaining space and imaginary time-dependent bosonic (phonon) degrees of freedom $x(r,\tau)$, using a weight which combines a boson action $S_{\rm B ose}$ and the fermion determinants (one for each fermionic species). If, as in the Hubbard model, quartic (interaction) terms in the fermions are present, they are decoupled via an
auxiliary field $h(r, \tau)$.  In either case, after the fermionic trace is performed, the sampling is now over these `classical' 
fields, and may be implemented by utilizing a variety of standard numerical techniques.

With the proliferation of computing resources over the last few decades, such QMC simulations have become indispensable tools for investigating difficult problems involving strong correlations in a variety of topics in condensed matter~\cite{ceperley95,foulkes01}, high energy~\cite{degrand06} and nuclear physics~\cite{carlson15}, as well as in chemistry~\cite{hammond94,needs20}, providing many breakthroughs in these fields. These successes notwithstanding, a pervasive problem afflicting such methods, limiting their scope of application considerably, is the sign problem (SP), which occurs when the fermion determinants become negative for certain quantum configurations, leading to a negative ``probability". 
This has given rise to a considerable body of research aimed at solving, or alleviating, the SP\cite{loh90,ortiz93,zhang97,chandrasekharan99,henelius00,bergkvist03,troyer05,nyfeler08,nomura14,mukherjee14,shinaoka15,kaul15,iglovikov15,he19,fukuma19,ulybyshev20,kim20}. Nevertheless, it remains unsolved, and one of the central issues in this regard is to understand how the underlying physics of the problem in consideration affects the SP. In the case of AFQMC, this is intimately related to the individual configurations of the bosonic or auxiliary fields and how they affect the sign of the fermion determinant.

The fermion determinant itself sums over all the possible quantum mechanical world lines of fermions moving in the instantaneous value of the physical bosonic field or the artificially introduced auxiliary field.  For a particular world line configuration, if the fermions wind around each other an odd number of times, the contribution to the determinant
is negative.  
This picture provides one view of
the origin of the sign problem:  
To the extent that different regions of the space-imaginary time lattice are uncorrelated, there is a constant `density'
of winding, and one expects that
as the spatial lattice size $N$ and inverse temperature $\beta$
grow, 
the likelihood of positive
and negative world line configurations becomes equal, and the
average sign vanishes exponentially with both $\beta$ and $N$.

The assumption that different regions are uncorrelated is non-trivial.  
One motivation is the absence of any intrinsic dynamics in the auxiliary field:
$h(r, \tau)$ couples to the fermionic degrees of freedom, but different components
of $h(r, \tau)$ are not coupled to each other.  Indeed, it is known that if 
$h(r, \tau)$ do interact the sign problem can be
mitigated~\cite{blankenbecler81,johnston03,bercx17}.  As an extreme example, 
if $h$ has no $\tau$ dependence, the sign of the determinant is positive.
Similarly, in electron-phonon models the phonon kinetic energy energy 
$\hat p^2/2m$ induces correlations in $x(r,\tau)$ on adjacent $\tau$
by penalizing imaginary time fluctuations, especially at low phonon frequencies $\omega$.
Although the details are complex, typically one expects the SP to be reduced. 

Symmetries might allow simulations to avoid the sign problem in
special situations~\cite{loh90,wu05,li16}.  The most simple scenario is one in which there
are two fermionic species (e.g.~spin up and spin down) which couple to the
auxiliary or bosonic field in the same way.  Then, as long as 
they also share common structure in the other pieces of the Hamiltonian
(the same hopping and chemical potential terms, for example) the two
matrices arising when the species are traced out are identical.  
Although the individual determinants can (and do) change sign, their
product is always a square, and hence positive.
Such situations are, however, not generic, and in most QMC, the sign problem
is significant.

This traditional argument~\cite{loh90} makes no explicit reference to
how the fields are selected, and hence suggests
that the average sign of the determinants of {\it randomly chosen} 
$h(r, \tau)$ 
should vanish.  
The intent of this paper is to investigate this issue further, and quantitatively.
Before doing so, it is useful to make some analogies with Monte Carlo for classical
degrees of freedom.

In classical statistical mechanics the expectation value of an observable
${\cal A}$ takes the form
\begin{align}
\langle {\cal A} \rangle & = {\cal Z}^{-1} \, {\rm Tr}_{\{h\}} 
\, {\cal A}(\{h\}) \, e^{-\beta E(\{h\})}
\nonumber
\\
{\cal Z} & =  {\rm Tr}_{\{h\}} \, e^{-\beta E(\{h\})}
\label{eq:classstatmech}
\end{align}
where $E(\{h\})$ is the energy of the system described by some collection
of degrees of freedom $\{h\}$ and $\beta$ is the inverse temperature.  
Implicit in the structure of Eq.~\ref{eq:classstatmech}
is that ${\cal A}$ does not depend on $\beta$.  However, certain ${\cal A}$
do have (trivial) $\beta$ dependence.  For example, in the paramagnetic
phase of the classical 
Ising model the magnetic susceptibility $\chi = \beta \sum_{ij} \langle S_i S_j \rangle$,
where $S_i=\pm 1$ are the Ising spins.  Similarly, energy-fluctuation-based measurements
of the specific heat measure the observable
$C = \beta^2 \big( \langle E^2 \rangle - \langle E \rangle^2 \big)$.
In such cases $\beta$ comes out
of the trace over the degrees of freedom.  Its presence does not affect the critical
properties.  In the case of the susceptibility, as one approaches $T_c$, the $\beta$ factor
merely multiplies the sum of the spin-spin correlations, $\langle \sum_{ij}S_i S_j\rangle$  by
$\beta_c$, but does not alter the power law describing its divergence.

In such a situation, a rescaling of the 
weight ${\cal W} = e^{-\beta E} \rightarrow ( e^{-\beta E})^g$ merely amounts to a shift in inverse temperature $\beta \rightarrow g \beta$ (or, equivalently, in the energy scales $E \rightarrow g E$) in the calculation of $\langle {\cal A} \rangle $. This analogy also makes clear that $g \neq 1$ will alter the expectation values measured, since it
changes the temperature of the simulation. As we shall see in detail below, in Quantum Monte Carlo (QMC) the situation is more complex. There, the observable $\hat {\cal A}$ can depend in a complicated way on the inverse temperature.  As a consequence, the change in  $\langle \hat {\cal A} \rangle $ upon a weight rescaling ${\cal W} \rightarrow {\cal W}^g$ can be highly non-trivial.

 The rescaling parameter $g$ has the effect of tuning the
configurations sampled
in a simulation.  $g=1$ gives the appropriate expectation values for the
energy (Hamiltonian) in question.  The limit $g=0$ makes all configurations of
$\{h\}$ equally likely.
In this paper we investigate the effect of tuning $g$ in determinant Quantum Monte Carlo (DQMC)~\cite{blankenbecler81}. Focusing on the average sign $\langle {\cal S} \rangle$, our purpose here is to understand whether the fermion SP in which $\langle {\cal S}\rangle \rightarrow 0$, occurs because the value of the sign itself for `random' configurations is becoming increasingly badly behaved as $\beta$ increases, as suggested by the winding argument above, or whether the sampling is preferentially guiding the system to a region in phase space where the weighted configurations yield vanishing average sign. To this end, we analyse three different realizations of the Hubbard model, viz, on the square lattice, on the honeycomb lattice and the ionic Hubbard model with a staggered potential, 
investigating the average sign as well as several physical variables, over a large range of parameter values such as the chemical potential $\mu$ and the interaction $U$, as we vary $g$ systematically. 

Our results show that the SP depends on the weight and temperature
in a non-trivial manner. In the square lattice model, we find that the SP originates mainly from the the fact that {\it weight guides the simulation to regions of phase space
with a low average sign}. In this situation, random HS fields result in a larger value of the overall sign at all temperatures and densities. 
In the honeycomb lattice, on the other hand, the SP becomes progressively worse for random sampling as the system approaches the interaction induced antiferromagnetic Mott insulator (AFMI), underscoring the complex relationship between the origin of the sign problem and the underlying physical states in DQMC. The ionic Hubbard model shows qualitatively similar results; here, the sign becomes worse again for random sampling as we navigate across the correlated metal (CM) phase and approach the AFMI state. In addition, analysis of various physical quantities as well as the detailed nature of the sign curves reveals that, in many respects, a reduced $g$ pushes the system to weaker coupling. Nevertheless, the system retains non-trivial signatures of the interaction in certain characteristics even in the fully random $g=0$ case. 

Next, we briefly consider the complementary `oversampling' case where $g>1$, for the square lattice model. In this case, signatures of a stronger effective coupling as $g$ is increased are even more prominent, as the Mott plateau around half-filling becomes progressively stronger, leading to a swift reduction in the SP accompanied by a shift in the sign minimum as $g$ is increased. We end with a quantitative analysis of the effective interaction with changing $g$, confirming the qualitative observations made earlier.


\section{Determinant QMC and the Hubbard Model}  
\label{sec:DQMCHM}

We investigate the sign problem within the context of the Hubbard Hamiltonian~\cite{fazekas, hubbard},
\begin{align}
\hat H &= \hat K + \hat V
\nonumber \\
\hat K &= 
-t \sum_{\langle i j \rangle} \big( \, 
c_{j \sigma}^{\dagger} c_{i \sigma}^{\phantom{\dagger}} + 
c_{i \sigma}^{\dagger} c_{j \sigma}^{\phantom{\dagger}}  \, \big)
- \sum_{i} \mu_i 
\big( \, n_{i \uparrow} + n_{i\downarrow} \, \big)
\nonumber \\
\hat V &= U \sum_{j} 
\big( n_{j \uparrow} - \frac{1}{2} \, \big)
\big( n_{j \downarrow} - \frac{1}{2} \, \big)
\end{align}
Here $c_{j \sigma}^{\dagger}
(c_{j \sigma}^{\phantom{\dagger}})$
are fermion creation(destruction) operators at spatial site $j$ and with spin $\sigma$.
$n_{j \sigma} = c_{j \sigma}^{\dagger} c_{j \sigma}^{\phantom{\dagger}}$
is the number operator.
$t$ is hopping matrix element between nearest neighbor sites,
$\mu$ is the chemical potential and $U$ is an on-site repulsion.
We set $t=1$ as our unit of energy.

As remarked earlier, we will explore several different contexts, beginning with the 2D square lattice
at constant $\mu_i = \mu$,
which is the most famous geometry owing to its relevance to
cuprate physics~\cite{scalapino94}.  We will then simulate the honeycomb lattice in order to understand if the effect of $g$ is linked in any way to the quantum critical point $U_c/t\simeq3.8$ which separates antiferromagnetic from semimetallic behavior~\cite{paiva05,meng10,sorella12,assaad13,Toldin2015,Otsuka2016}. Finally, we will turn to the ionic Hubbard model,  [$\mu_i = (-1)^i \, \mu$], on a square lattice to further test our conjectures.

We use a commonly employed DQMC formulation~\cite{blankenbecler81,white89}.  In computing
the partition function the inverse temperature is discretized,
$\beta = L_{\tau} \Delta \tau$, and the Trotter approximation is used to separate out the exponentials of the kinetic and potential energies. Thus, the partition function may be written as
\begin{align}
{\cal Z} &= {\rm Tr} ~( e^{- \beta \hat H}) \nonumber \\ &= {\rm Tr}~ \Big( e^{- \Delta \tau \hat H} \Big)^{L_{\tau}}
\nonumber \\
&\approx {\rm Tr}~ \Big(e^{- \Delta \tau \hat K(1)}~e^{- \Delta \tau \hat V(1)}~e^{- \Delta \tau \hat K(2)}~e^{- \Delta \tau \hat V(2)} \cdots \nonumber \\ 
&\cdots  e^{- \Delta \tau \hat K(L_{\tau})}~e^{- \Delta \tau \hat V(L_{\tau})}\Big)
\end{align}

The quartic on-site interaction term $\hat V$ is now decoupled by a discrete Hubbard Stratonovich (HS) transformation~\cite{hirsch83}:

\begin{align}
 e^{ - \Delta \tau U (n_{i \uparrow} - \frac{1}{2}) (n_{i \downarrow} - \frac{1}{2}) } &= {\cal C} \sum_{h_{i} = \pm 1} e^{\alpha h_{i} (n_{i \uparrow} - n_{i \downarrow}) },
\end{align}
where ${\rm cosh}(\alpha) = e^{ \Delta \tau U / 2}$, ${\cal C} = \frac{1}{2} 
e^{- \Delta \tau U / 4}$ and $h_{i}$ are discrete classical variables that
 only take values $\pm 1$. This converts a quartic fermionic term into a
 quadratic one, while adding a sum over the new variables $h_{i}$. Introducing
 this transformation for each lattice site $i$ at each time slice $l$,
 the partition function may be rewritten as

\begin{align}
 {\cal Z} = {\cal C}^{N L_{\tau}}~ {\rm Tr}_{\{h \}} ~{\rm Tr}~ & \Big( e^{ \Delta \tau \hat K}
 e^{{\hat {\cal V}} (1)}~e^{ \Delta \tau \hat K} e^{{\hat {\cal V}} (2)}
 \nonumber \\
  \cdots \, &e^{ \Delta \tau \hat K} e^{{\hat {\cal V}} (L_{\tau})} \Big),     
\end{align}
where $N$ denotes the number of lattice sites, the operator
${\hat {\cal V}}_{ij\sigma}(l) = c^{\dagger}_{i \sigma}~ v^{ij}_{\sigma}(l) ~ c_{j \sigma}$,
 with $v^{ij}_{\sigma}(l) = \alpha \, \sigma h_{i}(l) \, \delta_{ij}$, where $\sigma = \pm 1$, 
 and $\hat K_{ij\sigma}(l)
 =  c^{\dagger}_{i \sigma}~ k^{ij}_{\sigma}(l) ~ c_{j \sigma}$, $k^{ij}_{\sigma}(l)
 = (t) \delta_{\langle ij \rangle} +\mu_i \delta_{ij}$. For a given configuration 
$\{ h \}$, the terms in the exponentials are all quadratic, allowing us
 to perform the fermionic trace analytically, yielding~\cite{blankenbecler81,white89}  
\begin{align}
 {\cal Z} = {\cal C}^{N L_{\tau}}~ {\rm Tr}_{ \{ h \} }~~ [{\rm det} M_{\uparrow}(h)]~ [{\rm det} M_{\downarrow}(h)],     
\end{align}
where the matrix $M_{\sigma}(h) = (I + \prod_{l} e^{\Delta \tau k} e^{v_{\sigma}(l)}) $.

All physical observables can be expressed in terms of the fermion
Green's function ${\cal G}_{\sigma, ij} = \langle c^{\phantom{\dagger}}_{i\sigma}
c^\dagger_{j\sigma} \rangle = 
M^{-1}_{\sigma i j}$.  For example, the fermion density
on site $i$ with spin $\sigma$, 
$n_{i\sigma} = 1 - {\cal G}_{\sigma, ii}$, the kinetic energy
(excluding the chemical potential)
$= \langle {\cal K} \rangle = (8t) {\cal G}_{i\, i+\hat x, \sigma}$
where the factor of eight arises from the two spin species, the two directions
$x,y$ to hop, and the Hermitian conjugate pair associated with hopping
$i \leftrightarrow j$.  Finally, the pair correlator
$P^d_{ij}= \langle \Delta^{d} _i \Delta^{d \dagger}_j \rangle$, where
$\Delta^{d \, \dagger}_j 
= c_{j \uparrow}^{\dagger} \big( \,\,
c_{j+\hat x \downarrow}^{\dagger} 
-c_{j+\hat y \downarrow}^{\dagger} 
+c_{j-\hat x \downarrow}^{\dagger} 
- c_{j-\hat y \downarrow}^{\dagger}  \big)
$ for $d$-wave symmetry.


It is important for us to re-emphasize the goal of this paper in introducing the
parameter $g$.  We are {\it not} seeking to find an improved importance
sampling scheme which would reduce the sign problem and associated error bars while
leaving physical observables with the same expectation values.
Such work has been productively undertaken by several groups in the community, especially within the context of `constrained path' auxiliary field QMC
\cite{carlson99,zhang95,zhang03,purwanto04,motta14,motta18}.
Instead, our objective is solely focussed on gaining insight into the origin of 
{\it sign problem itself}, and isolating whether it
is better or worse for {\it randomly selected} field configurations compared to
those chosen preferentially according to the weight (fermion determinants).

\section{Weak and Strong Coupling Limits}  \label{sec:WSCLimits}

In the $U=0$ limit, the matrices $M_\sigma$
and their inverses, the Green's functions ${\cal G}_\sigma$, are independent of the HS field configuration. As a consequence, expectation values of any observable $\hat A$ are exact and also independent of $g$. The proof is straightforward:
\begin{align}
\langle \hat A \rangle &=  {\cal Z}^{-1}
\int {\cal D}h \, \hat A(h) \,
\big( {\rm det} M_{\uparrow}(h)
{\rm det} M_{\downarrow}(h) \big)^g
\nonumber \\
{\cal Z} &=
\int {\cal D}h \,
\big( {\rm det} M_{\uparrow}(h)
{\rm det} M_{\downarrow}(h) \big)^g
\nonumber \\
\Rightarrow
\langle \hat A \rangle &=  
\frac{ \int {\cal D}h \, \hat A(h) \,  }
{ \int {\cal D}h }
\end{align}

Here, the notation $\int {\cal D}h$ is used as a general symbol for summation or integration over the degrees of freedom $h$, including both the cases where they represent continuous variables (for a continuous HS transformation) and discrete ones (as in our calculations below).

The strong coupling (single site) limit $t=0$ is more interesting. In this case, the sites are completely decoupled and the solution reduces to a product of single site calculations. The partition function for a general value of $g$ is given by

\begin{align}
 {\cal Z} &= {\cal C}^{N L_{\tau}}~\prod_{i} \sum_{ \{ h_{i} \}}  \Big\{ \prod_{\sigma} \Big( 1 + e^{\beta \mu_i + \alpha \sigma \sum_{l} h_{i}(l)} \Big) \Big\}^{g} \end{align}

This can be evaluated analytically at $g=1$ (and trivially at $g=0$). 
The general case with $g \neq 1$ 
can be easily evaluated numerically. In Appendix~\ref{sec:app5}, we show detailed results for the number density $\langle n \rangle$ and the double occupancy $\langle n_{\uparrow} n_{\downarrow} \rangle$ for several values of $g$, $U$ and $\beta$. These results show that this simple case already demonstrates aspects of the non trivial effects of varying the nature of
the high probability configurations in such problems. 
\section{Hubbard Model on a Square Lattice}  \label{sec:SquareLattice}

\begin{figure}[t]
\includegraphics[scale=0.27]{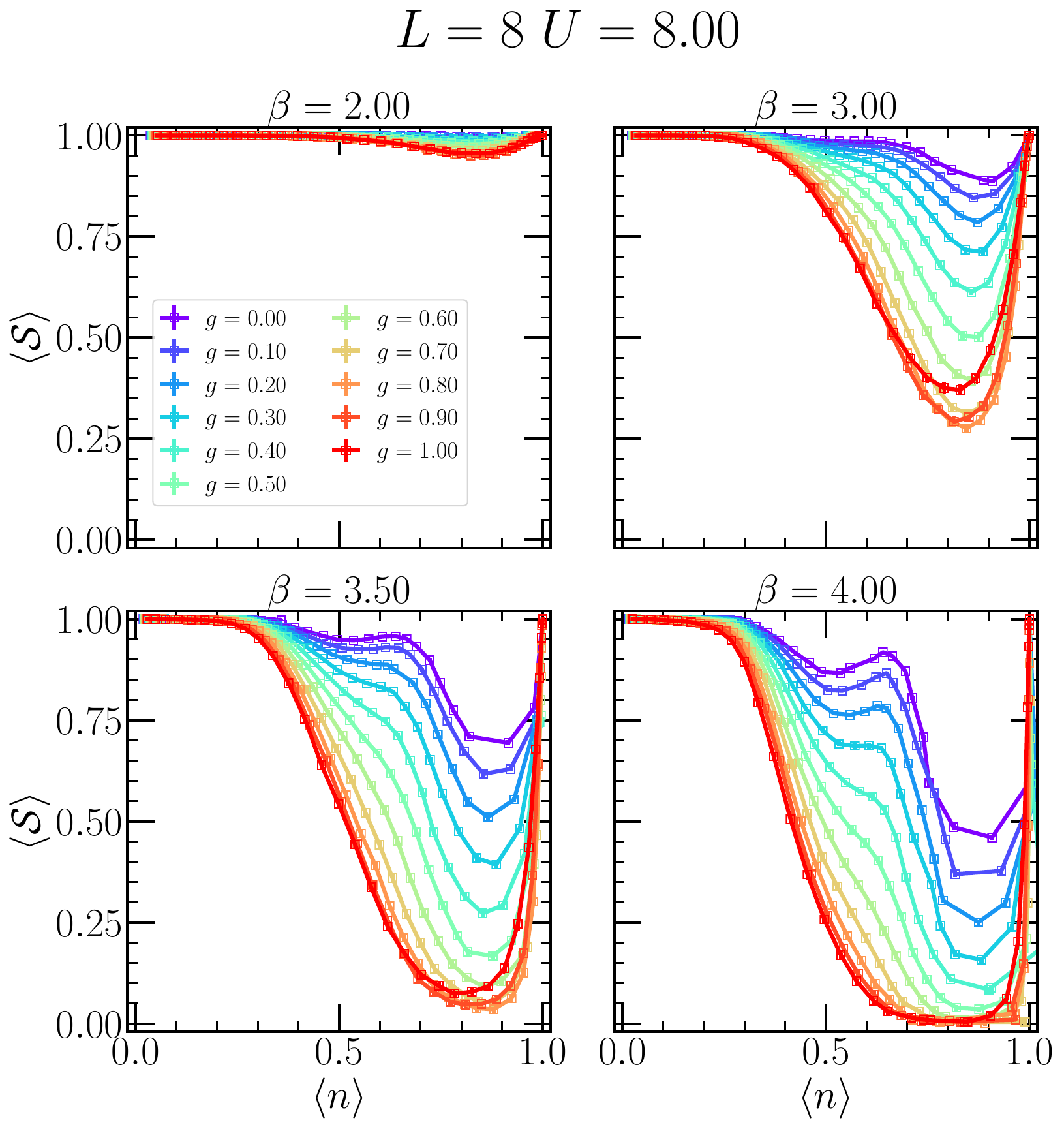}
\caption{The average sign $\langle {\cal S} \rangle$ vs.~$\langle n \rangle$ at $U=8$ as a function of $g$ on an $8 \times 8$ lattice for four values of $\beta$. As $\beta$ increases, the minimum in $\langle {\cal S} \rangle$ becomes wider and deeper, as expected, displaying a minimum around $\langle n \rangle_{c} \sim 0.85$. While the sign remains close to $1$ for $g \rightarrow 0$ at $\beta \lesssim 3$, it starts reducing noticeably as $\beta$ is increased further, eventually forming a `valley-peak-valley' structure as seen in the bottom panels. This suggests that while the bulk of the sign problem at low temperature originates from the restricted configuration space available to the system, random configurations of the HS fields also result in a perceptible reduction in $\langle {\cal S} \rangle$ at low enough temperatures. The bottom panels, where the sign is the worst, show a sharp upturn near half-filling, since the total sign is constrained to be equal to $1$ at $\mu = 0$ due to particle-hole symmetry. 
}
\label{fig:fig1}
\end{figure}

In this section, we consider the Hubbard model on a square lattice of size $L$ with a constant $\mu_{i} = \mu$. As is well known, at half filling, the on-site interaction $U$ results in an AFMI. Away from half filling, the possibility of superconducting correlations mediated by spin fluctuations makes this iconic Hamiltonian relevant as a prospective model for high temperature superconductors~\cite{scalapino94}. Thus, it provides a rich background for investigating the effect of the rescaling parameter $g$ on the evolution of the system in various parameter regimes.  

Figure~\ref{fig:fig1} shows the average of the total sign $\langle {\cal S} \rangle$ vs. the total density $\langle n \rangle$ for different values of the rescaling parameter $g$ at four different values of $\beta$. For $\beta \leq 3$ (top panels), it is seen that as $g \rightarrow 0$,
$\langle S \rangle \rightarrow 1$, with only a small dip around $\langle n \rangle \sim 0.8$. These plots suggest that at temperatures that are not too low, simulations which sample with randomly chosen HS field configurations tend to have a considerably reduced SP. On the other hand, at lower temperatures, $\beta \gtrsim 3$ (lower panels), we find that while the sign continues improving systematically as $g$ is reduced from $1$ towards $0$, the value of $\langle {\cal S} \rangle$ around its minimum is decreased considerably compared to larger temperatures even near $g=0$. In addition, we find an emerging valley-peak-valley structure as $g \rightarrow 0$, reminiscent of the shell effect in non-interacting finite size lattices. In Appendix~\ref{sec:app4}, we explore this connection more carefully by considering even lower temperatures, and demonstrate that the values of $\langle n \rangle$ at the maxima of the sign correspond to the locations of the density steps on a non-interacting lattice of the same size due to the shell effect~\cite{Mondaini2012}. An additional feature of the data, seen at all temperature values presented here, is that there is an initial {\it worsening} of the SP for $0.8 \lesssim g < 1$. 





\begin{figure}[b]
\includegraphics[scale=0.27]{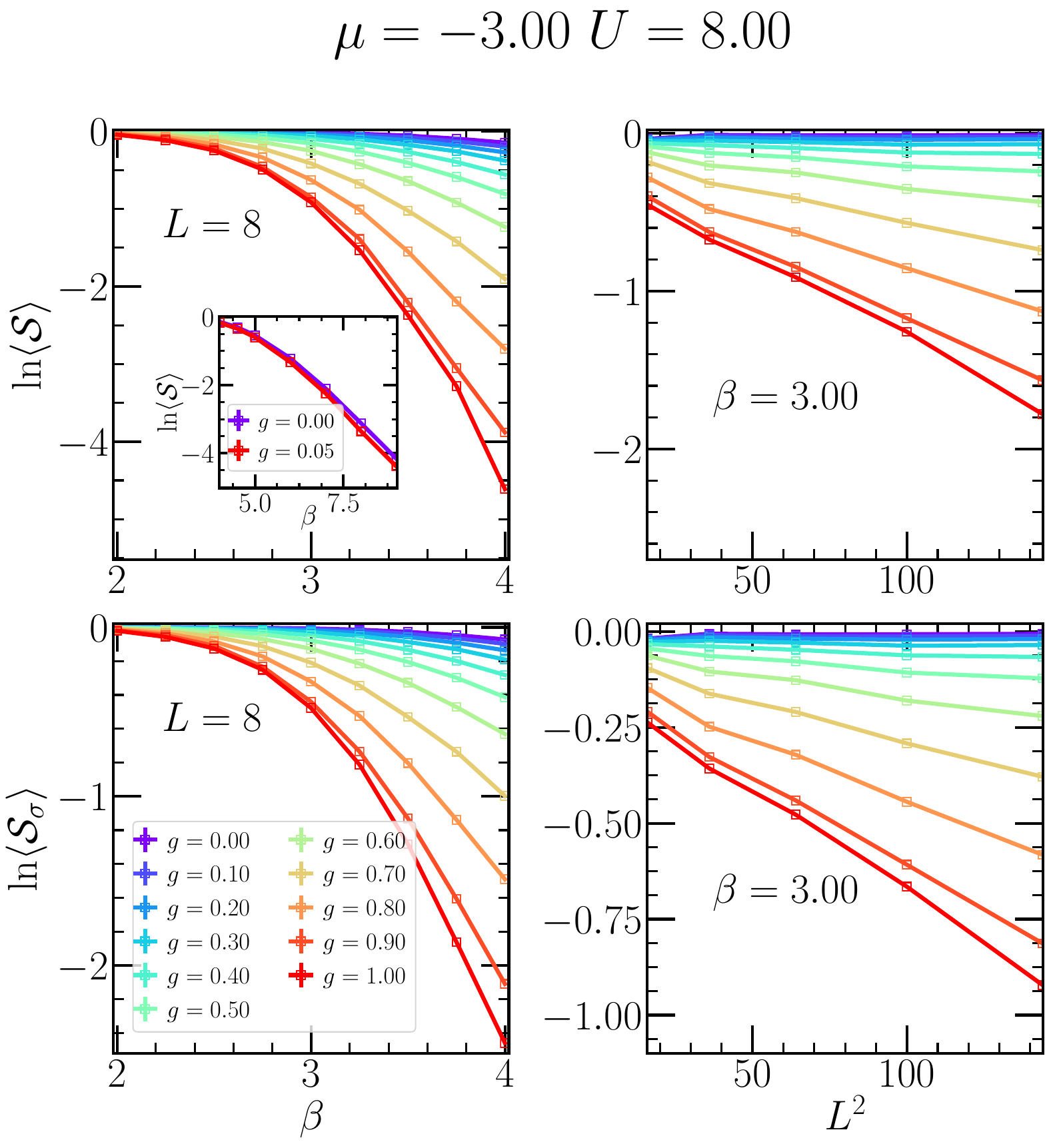}
\caption{Scaling of the total and spin resolved sign with temperature and system size at $\mu = -3.0$, $U=8$ on an $8 \times 8$ lattice. Left column shows ${\rm ln}(\langle {\cal S} \rangle)$ and ${\rm ln}(\langle {\cal S}_{\sigma} \rangle)$ vs. the inverse temperature $\beta$ for a range of values of $g$. We find a faster than exponential reduction in the sign with decreasing temperature, both for the total and spin resolved quantities. The inset in the top panel extends this to temperatures $\beta \sim 9$ for $g = 0.0$ and $0.05$, demonstrating that even the randomly sampled case shows a faster than exponential scaling over a large temperature range. The right column plots the same quantities as a function of the lattice size, $L^{2}$. The scaling in this case is exponential for all $g$ values.
}
\label{fig:fig2}
\end{figure}

As explained earlier, the traditional argument about the connection of the sign with the fermionic world lines suggests that with lowering temperature (i.e., increasing $\beta$) and increasing system size, the fermion world lines are more likely to wind around each other and provide determinants of both sign more frequently, resulting in a progressively worse SP on the average. In order to see how the choice of $g$ affects the scaling properties of the sign, we show, in Fig.~\ref{fig:fig2}, the scaling properties of $\langle {\cal S} \rangle$ as well as the spin resolved sign, $\langle {\cal S}_{\sigma} \rangle = (\langle {\cal S}_{\uparrow} \rangle + \langle {\cal S}_{\downarrow} \rangle) / 2$, with increasing lattice size $N=L^{2}$, and inverse temperature $\beta$. We find that, for $g \lesssim 1$, where the HS fields are sampled according to the correct thermal weights, both show a faster than exponential drop with increasing $\beta$, as found in earlier work~\cite{iglovikov15}. As $g$ is reduced, the overall sign gradually improves at all $\beta$ values, but the reduction with increasing $\beta$ remains faster than exponential. The plots near $g=0$ seem to show little change in this temperature range, but, as the inset demonstrates, at smaller temperatures, they show the same faster than exponential reduction. The scaling with system size at fixed temperature, as seen in the right column, is exponential. Again, we find that while $\langle {\cal S} \rangle$ is less severe at lower $g$ values, it continues to reduce exponentially with increasing lattice size. 

\begin{figure}[t]
\includegraphics[scale=0.22]{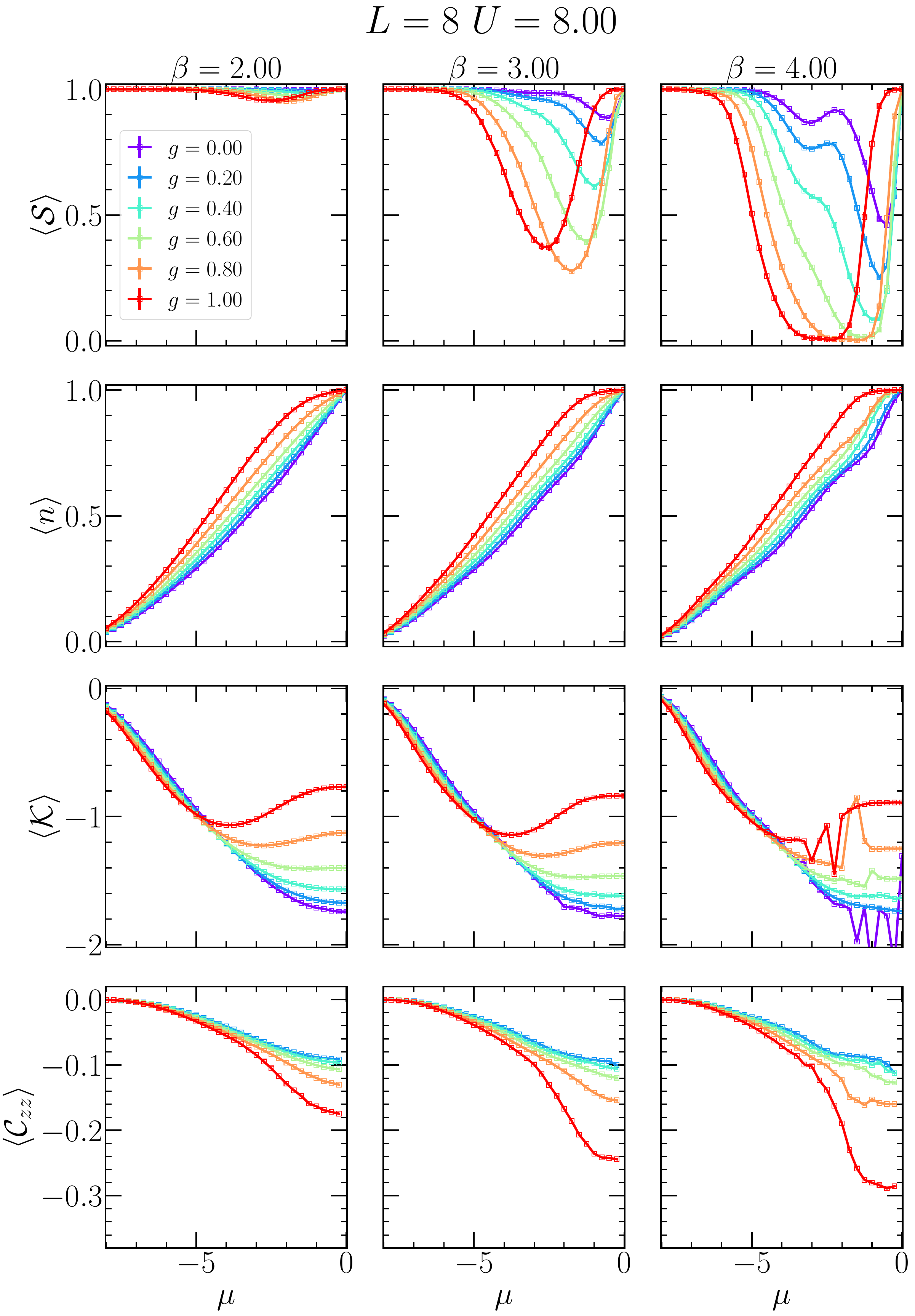}
\caption{Plots of the average sign $\langle {\cal S} \rangle$ (first row) and observables including the density $\langle n \rangle$ (second row), kinetic energy $\langle {\cal K} \rangle$ (third row) and  nearest neighbor spin-spin correlator $C_{zz}$ (fourth row), vs $\mu$ as a function of $g$ for three different values of $\beta$. These physical observables clearly demonstrate signatures of a reduced effective interaction (see text for more details).
}            
\label{fig:fig3}
\end{figure}

These results indicate that while the SP (both the total value as well as the spin resolved ones) itself is dominated by the reduced phase space at lower temperatures and larger lattice sizes, the scaling dependence on both parameters remains very similar throughout the whole range of the sampling parameter $g$.

Near $g=1$, the thermodynamic sampling restricts the HS field configurations to a suitably confined region of the total phase space, reflecting the underlying physics, especially at low temperatures. As $g$ is reduced from $1$, the increasingly random sampling occupies progressively larger regions of phase space, affecting the physical properties of the system. In order to study this more carefully, we show a number of physical variables including the number density $\langle n \rangle \equiv \frac{1}{N} \sum_{i,\sigma} \langle n_{i\sigma}\rangle$, the kinetic energy $\langle {\cal K} \rangle$ and the nearest neighbour spin-spin correlations, $C_{zz} \equiv \frac{1}{4} \langle (n_{i\uparrow} - n_{i\downarrow})(n_{j,\uparrow} - n_{j,\downarrow})\rangle$ ($i,j$ are nearest neighbors),
as a function of the chemical potential $\mu$ in Fig.~\ref{fig:fig3}, for three temperature values. 


The second row demonstrates some of this physics for $g=1$ by plotting $\langle n \rangle$ vs. $\mu$. As the temperature decreases, the number density saturates close to half filling as the AFMI sets in and the compressibility goes to zero, triggering a Mott plateau. As $g$ is reduced and the HS fields are sampled more and more randomly, we find that this physics disappears slowly; the plots near $g=0$ do not show any signs of saturating. Instead, we find gentle oscillations in $\langle n \rangle$, reminiscent of the oscillations we saw in the sign for similar parameter values in Fig.~\ref{fig:fig1}. As mentioned above, we demonstrate in Appendix~\ref{sec:app4} that this is due to the re-emergence of finite size effects from the non-interacting problem. 

In the third row, we show the kinetic energy, $\langle {\cal K} \rangle$, for the same parameter values. Again, we see that while the $g=1$ results demonstrate the expected reduction in the kinetic energy as the density increases and the electrons start to avoid each other due to the on-site coupling $U$, lower values of the sampling parameter nullify this effect. We emphasize, however, that even at $g=0$, the system is very different from a non-interacting one in many respects, as seen, for instance, in the enlarged bandwidths in these results. 

The fourth row, which plots $C_{zz}$, also demonstrates the same suppression of the physics with lowering $g$. At $g=1$, as the system approaches half filling and the Mott state begins to appear, the nearest neighbour spin correlations become negative, their magnitude increasing with lowering temperature as the antiferromagnetic order becomes stronger. The weakening Mott insulator that results as $g$ is lowered leads to a corresponding reduction in the spin correlations, as is clearly evident from the plots.

Hence, we see that as $g$ is reduced, the physical variables clearly demonstrate signatures of a reduced effective interaction in some aspects, while retaining non-trivial signatures of the full interaction in others. This is not unreasonable, as the HS fields act as proxies for the electron-electron interactions~\cite{hirsch83}, and randomizing them progressively is expected to dilute the effect of $U$. On the other hand, the matrix elements in the simulations still contain exponential factors that include the full interaction strength, and, hence, certain features, such as the bandwidth, essentially retain their $g=1$ values throughout. In Section~\ref{sec:renorm}, we will quantify this relationship between the effective interaction $U_{\rm{eff}}(g)$ and $U$ by analysing physical observables such as the double occupancy $\langle n_{\uparrow} n_{\downarrow} \rangle$ and the kinetic energy $\langle \mathcal{K} \rangle$ in an attempt to put these observations on a firmer footing.

\begin{figure}[h]
\includegraphics[scale=0.27]{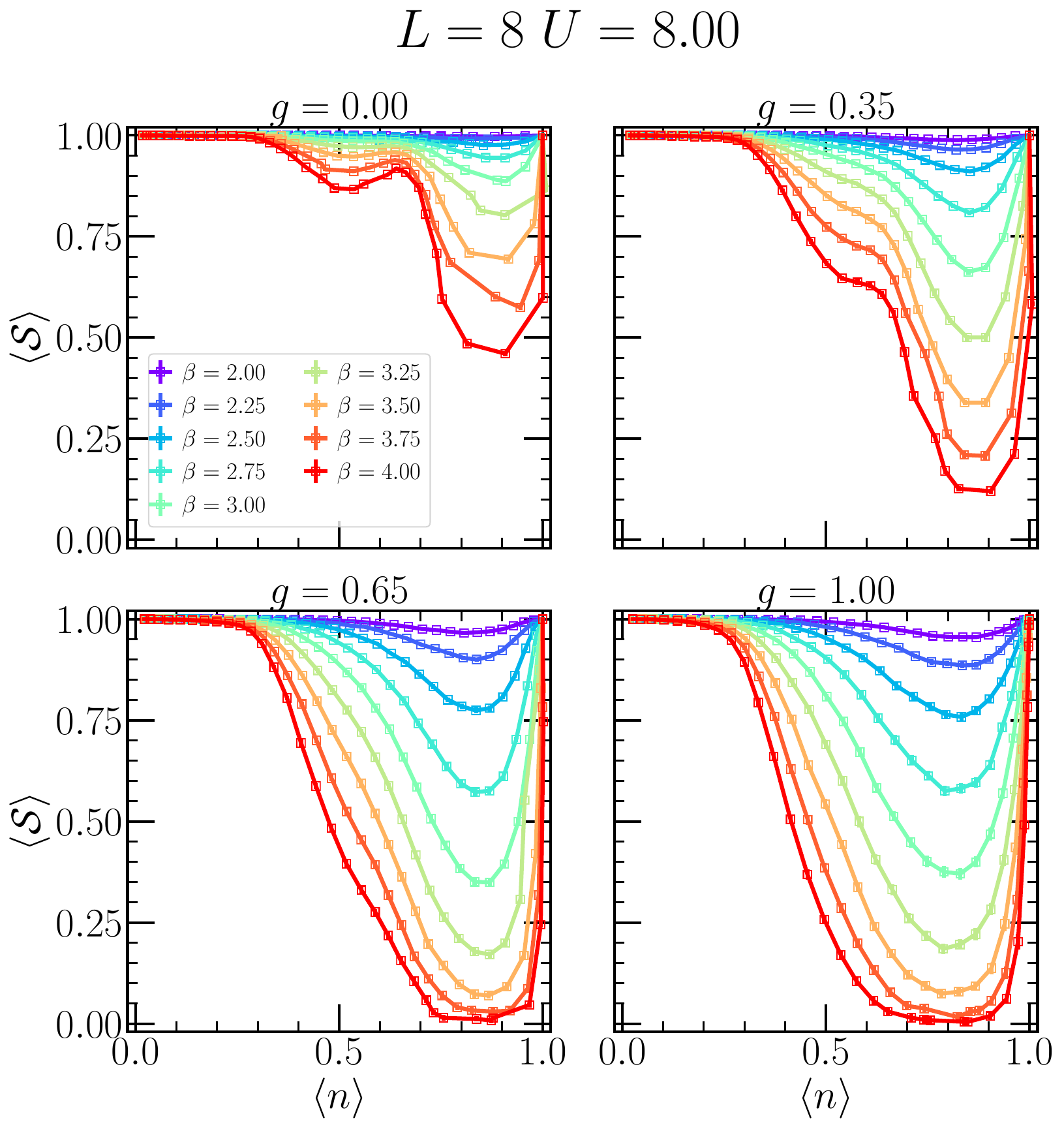}
\caption{The average sign $\langle {\cal S} \rangle$ vs. $\langle n \rangle$ at $U=8$ on an $8 \times 8$ lattice as a function of $\beta$ for four different $g$ values. Upper panels for $g < 0.4$ demonstrate finite size oscillations at lower temperatures. 
}
\label{fig:fig4}
\end{figure}

Now, we return to the behaviour of the sign again, focusing on the temperature dependence at fixed $g$ as well as the $g$ dependence at fixed $\mu$.  
In Fig.~\ref{fig:fig4}, we concentrate on the temperature dependence of the sign at fixed representative values of $g$. As expected, we find that the sign worsens with decreasing temperature for all $g$ values. In the lower panels, for larger values of $g$, the plots are smooth and rather similar. However, the top panels, with $g < 0.4$ show clear evidence of finite size based oscillations at lower temperatures, as remarked earlier in this paper.

\begin{figure}[h]
\includegraphics[scale=0.27]{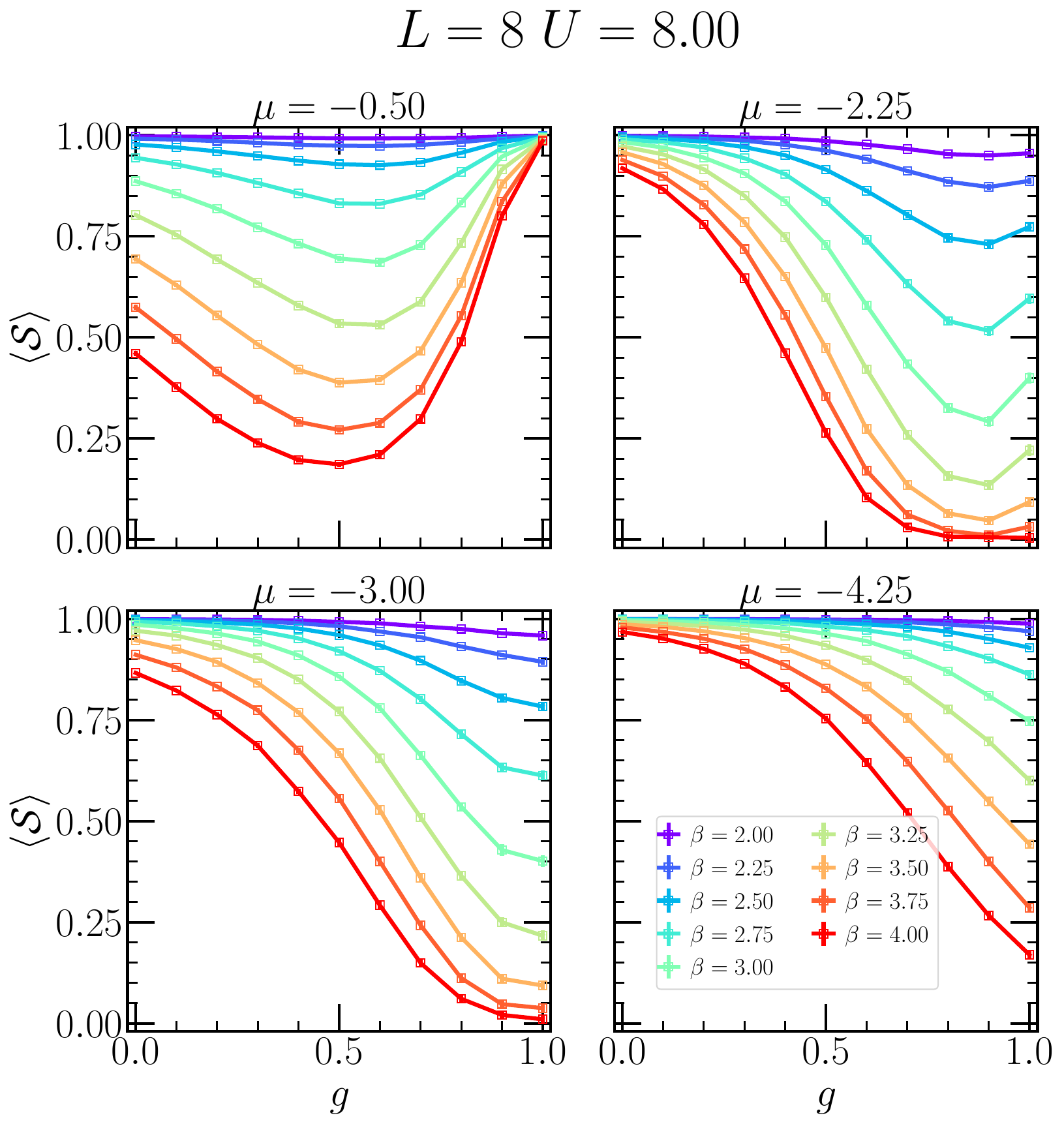}
\caption{The average sign $\langle {\cal S} \rangle$ vs. $g$ at $L=8$, $U=8$ as a function of $\beta$ for four different $\mu$ values. 
}
\label{fig:fig5}
\end{figure}

In Fig.~\ref{fig:fig5}, we replot the data with the sign $\langle {\cal S} \rangle$ as a function of the parameter $g$ with temperature at different values of $\mu$. As the top right and lower panels, with $\mu \lesssim -2.0$ show, increasing $g$ leads to a worsening of the sign at all temperatures. In contrast, the top left panel, with $\mu = -0.5$, close to half filling, shows a non-monotonic behaviour where the sign first reduces, reaches a minimum at a moderate value, and then increases continuously to unity as $g \rightarrow 1$, as the approaching Mott insulating state pins the density $\langle n \rangle \sim 1$, establishing particle-hole symmetry and mitigating the sign problem.

Thus, we find that on the whole, the SP in the standard square lattice Hubbard model is mainly due to the fermion determinants steering the simulation
to regions of low average sign. Random sampling typically mitigates the problem substantially over a large parameter window, even though it reappears at sufficiently low temperatures. The random fields tend to reduce the effect of the interaction, pushing the system to weaker coupling, resulting in a re-emergence of certain non-interacting features such as the shell effect, as well as reducing the signatures of the AFMI in observables such as the density, the kinetic energy and the spin-spin correlations. Even so, this comparison only holds for certain characteristics, and such systems retain several non-trivial signatures of interactions even at $g=0$. 


\section{Hubbard Model on a Honeycomb Lattice}  \label{sec:HCH}

\begin{figure}[t]
\includegraphics[scale=0.27]{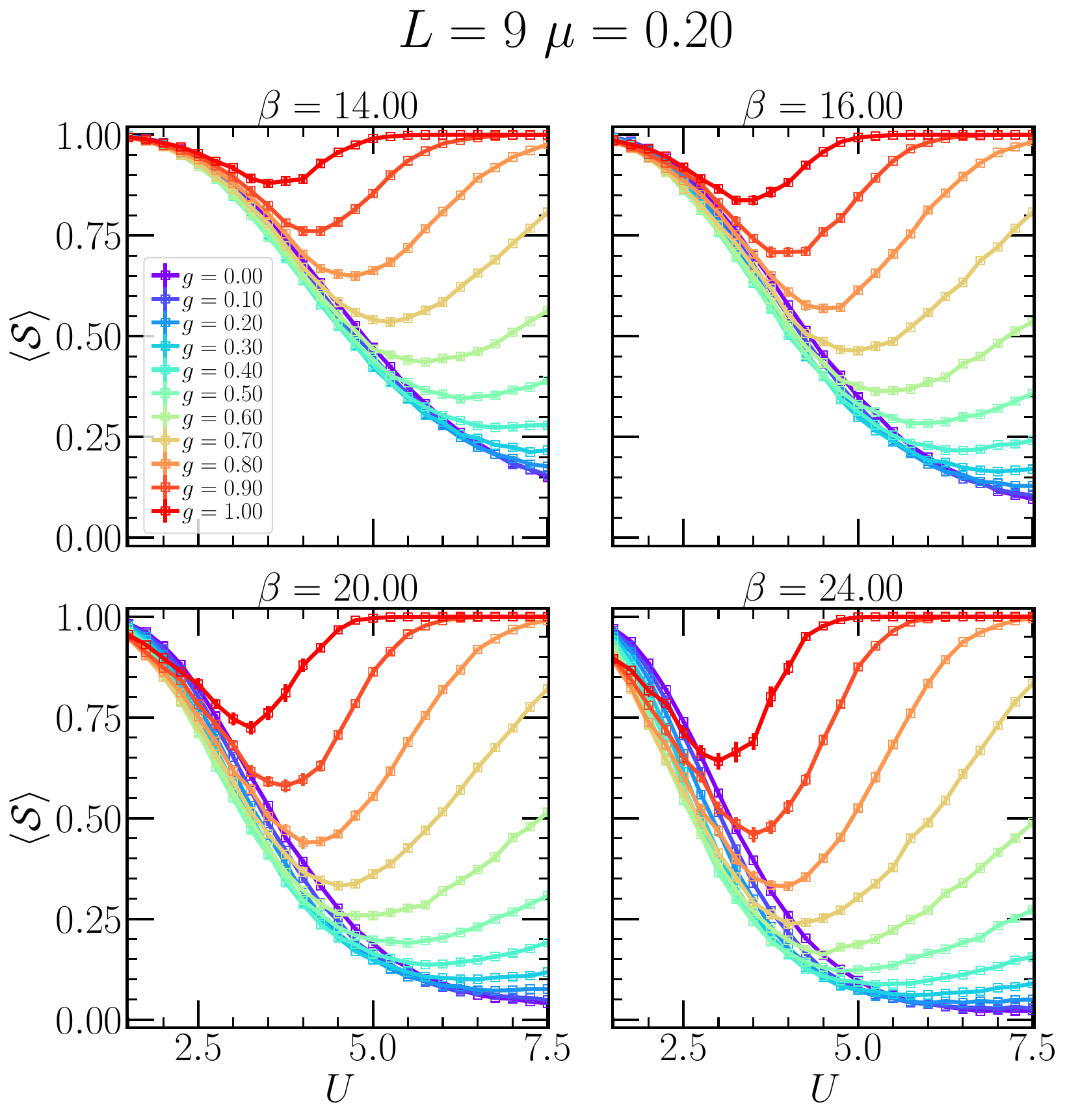}
\caption{The average sign $\langle {\cal S} \rangle$ vs. $U$ at $\mu = 0.2$, for the Hubbard model on a honeycomb lattice with $L=9$, with varying $U$ as a function of $g$, for four values of $\beta$. At $g=1$, $\langle {\cal S} \rangle$ shows a minimum around $U_{c}$. Reducing $g$ lowers the sign, shifting the minimum to the right (see text for details).
}
\label{fig:fig6}
\end{figure}


In the previous section, we analysed how the rescaling parameter $g$ affected the sign problem and physical variables of the square lattice Hubbard Hamiltonian. While this model has a very rich phenomenology, the Hubbard model on the honeycomb lattice provides added insight into the sampling problem by presenting a sharply defined quantum phase transition. 

At $U=0$, the Hubbard model on a honeycomb lattice is a semi-metal whose energy $\epsilon(k)$ disperses linearly with the momentum $k$ close to special points on the k-space called Dirac points. This leads to a semi-metallic density of states which also varies linearly with energy $\omega$ near $\omega=0$. As $U$ is turned on, the semi-metal state persists up to a critical coupling value $U=U_{c} \sim 3.8$~\cite{paiva05,meng10,sorella12,assaad13,Toldin2015,Otsuka2016}, where it undergoes a quantum phase transition into an AFMI, unlike the square lattice version, which shows AF order at any non-zero value of $U$.

In Fig.~\ref{fig:fig6}, we plot the average sign $\langle {\cal S} \rangle$ on a lattice with $L=9$ at a small non-zero $\mu = 0.2$ (the model is sign problem free at half filling) vs. the coupling $U$ as a function of $g$ for four different values of the temperature. The $g=1$ plots deviate from their limiting values of $\langle {\cal S} \rangle \rightarrow 1$ in a broad region around $U_{c}$, tracking the transition, as we have shown in detail elsewhere~\cite{sign-prob1}. As $g$ is reduced, we find that the dip in the sign becomes more pronounced, in stark contrast to the results of the square lattice Hubbard model where this led to a reduction in the sign problem, and the minimum shifts to higher values of $U$. An intuitive explanation of the latter result follows from the observation we have already made earlier: as the sampling becomes more random, the effect of $U$ is reduced. As a result, the semi-metal to AFMI transition here is pushed to higher values of $U$. This does not lead to a reduction in the sign, however, as full inclusion of the fermion determinants ($g=1$) in this case evidently leads the system to a phase space region (corresponding to the AFMI phase) where the sign problem is less severe, compared to $g=0$, where the determinants are ignored.  Since the AFMI phase is characterized by a finite regime where $\langle n \rangle \sim 1$ is saturated at $1$, reinstating particle hole symmetry (originally broken by the small but non-zero $\mu$), the sign may be expected to be robust in this phase.

Overall, these results again demonstrate the complex effect the underlying physics of the system can have on the SP in DQMC.

\begin{figure}[b]
\includegraphics[scale=0.27]{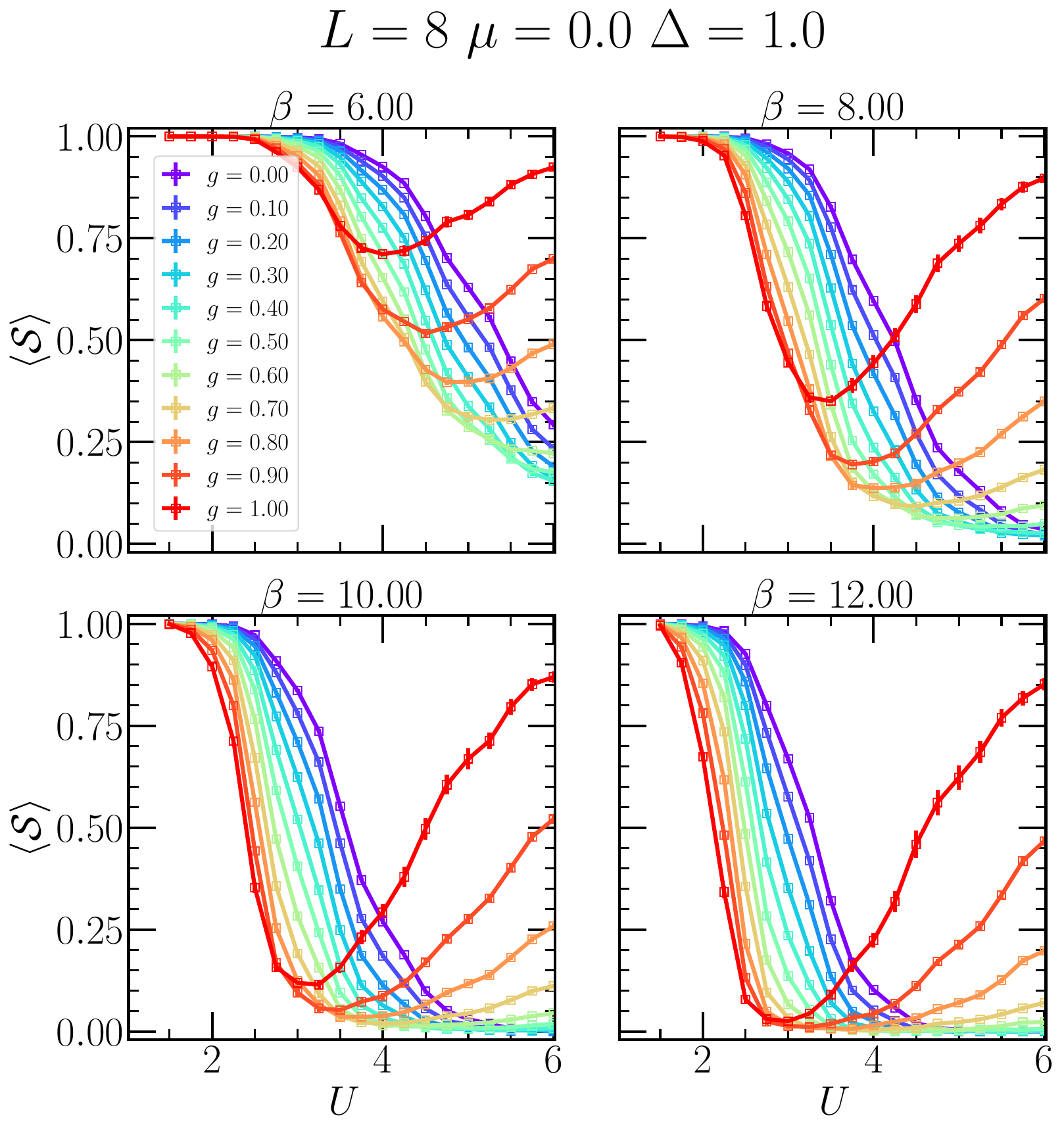}
\caption{The average sign $\langle {\cal S} \rangle$ vs.~$U$ at half filling $\mu = 0$, as a function of $g$ on an $8 \times 8$ lattice for four values of $\beta$ in the ionic Hubbard model. The staggered potential $\Delta = 1.0$. Results are qualitatively similar to the honeycomb model (see text).
}
\label{fig:fig7}
\end{figure}

\section{Ionic Hubbard Model}  \label{sec:BIAF}


In this section, we shift our attention to the ionic Hubbard model at half filling. On a square lattice, this model essentially consists of an added local staggered potential $\mu_{i} = (-1)^{i} \Delta / 2$ to the square lattice Hubbard model. In the non interacting limit $U / \Delta \rightarrow 0$, the system is a band insulator (BI) where the sites with a negative value of the potential, $-\Delta/2$, have a higher occupancy than the sites with a positive value, resulting in a charge density wave order due to the breaking of the sub-lattice symmetry by the staggered potential. In the opposite limit $U/ \Delta \gg 1$, the system is an AFMI due to the large onsite repulsion favouring single occupancy everywhere. Interestingly, as the coupling strength $U$ is increased starting from the weak coupling limit, the BI does not undergo a direct transition to the AFMI. Instead, over a range of values of $U$ and $\Delta$ where the two energy scales are comparable, the model displays an exotic correlated metal (CM) phase, as past work has demonstrated~\cite{paris07,chattopadhyay19}. 

Fig.~\ref{fig:fig7} plots the sign $\langle {\cal S} \rangle$ vs. the coupling $U$ at different $g$ values for four values of $\beta$, as before. At $g=1$, the sign shows a broad minimum roughly corresponding to the CM phase (the correspondence becomes sharper at lower temperatures). Similar to the results for the honeycomb Hubbard model, we find that increasingly random sampling worsens the SP and shifts the minimum value of the sign to the right, leading to a complete flattening of the curves (at least in the coupling range shown here) for $g \lesssim 0.5$ as the AFMI state becomes progressively weaker. The shift in the minimum suggests that the BI-CM boundary is also pushed to higher values of $U$, consistent with the reduction in the effective interaction.

\begin{figure}[t]
\includegraphics[scale=0.135]{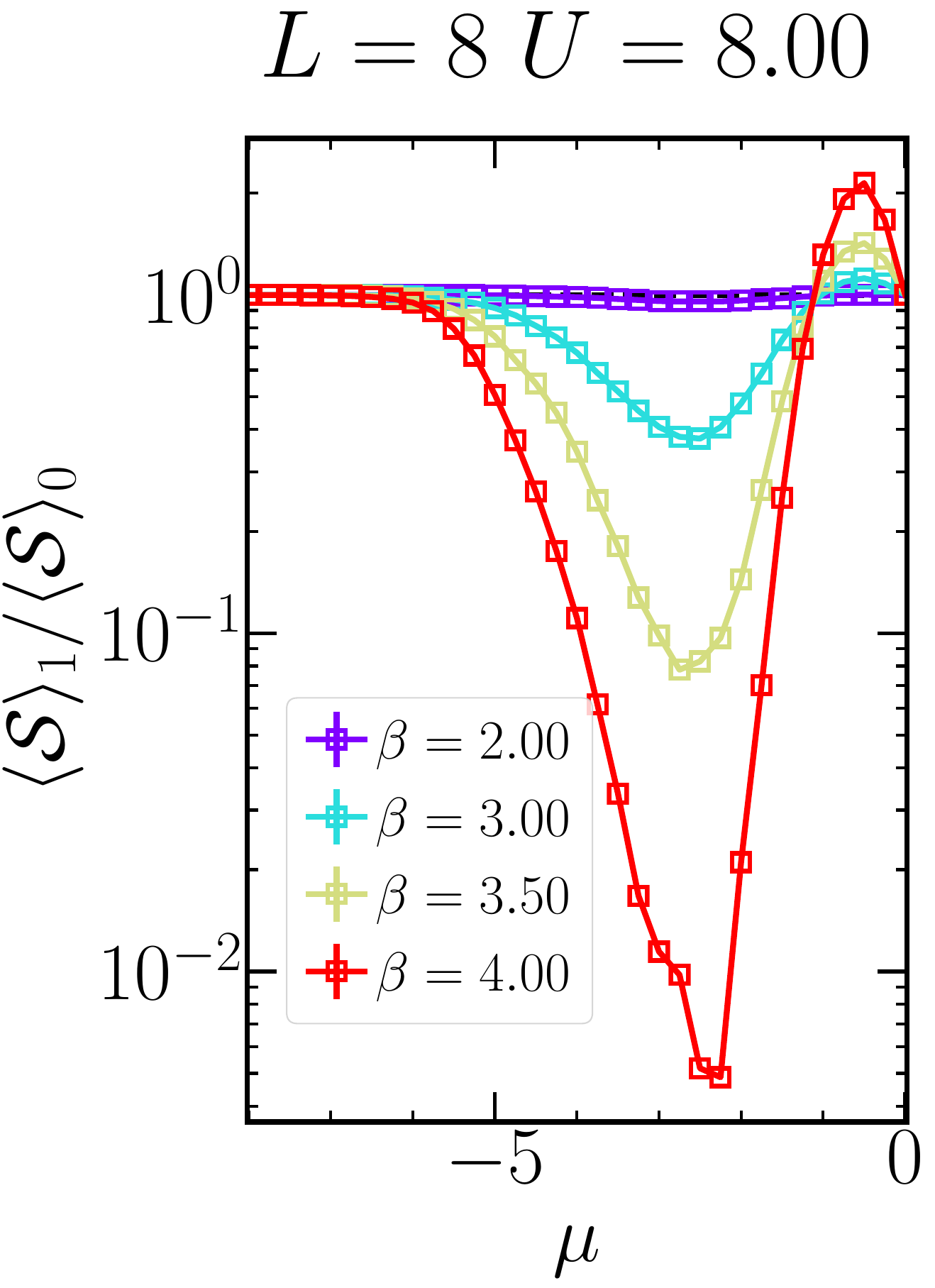}
\hspace{-0.2cm}
\includegraphics[scale=0.135]{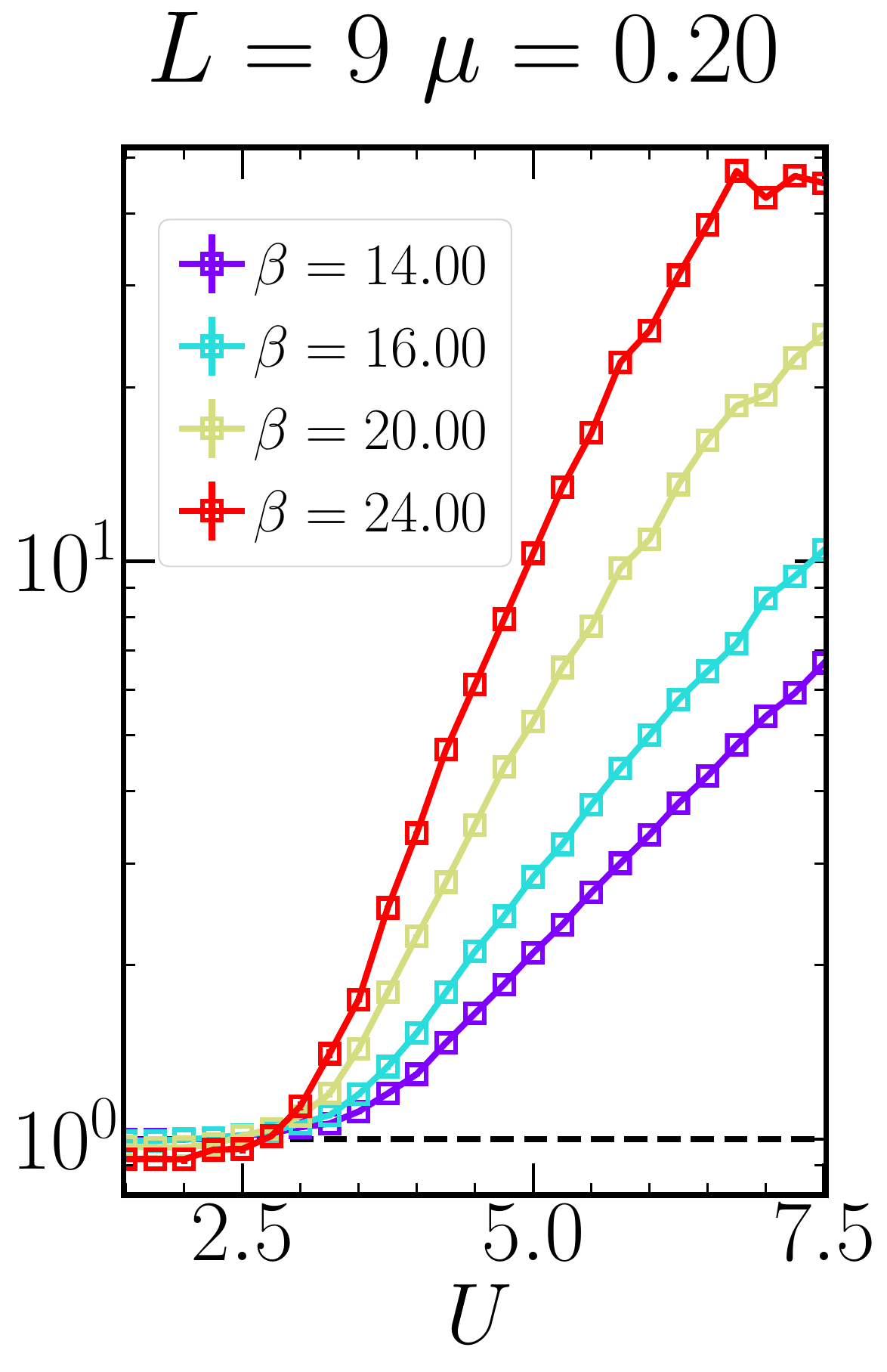}
\hspace{-0.2cm}
\includegraphics[scale=0.135]{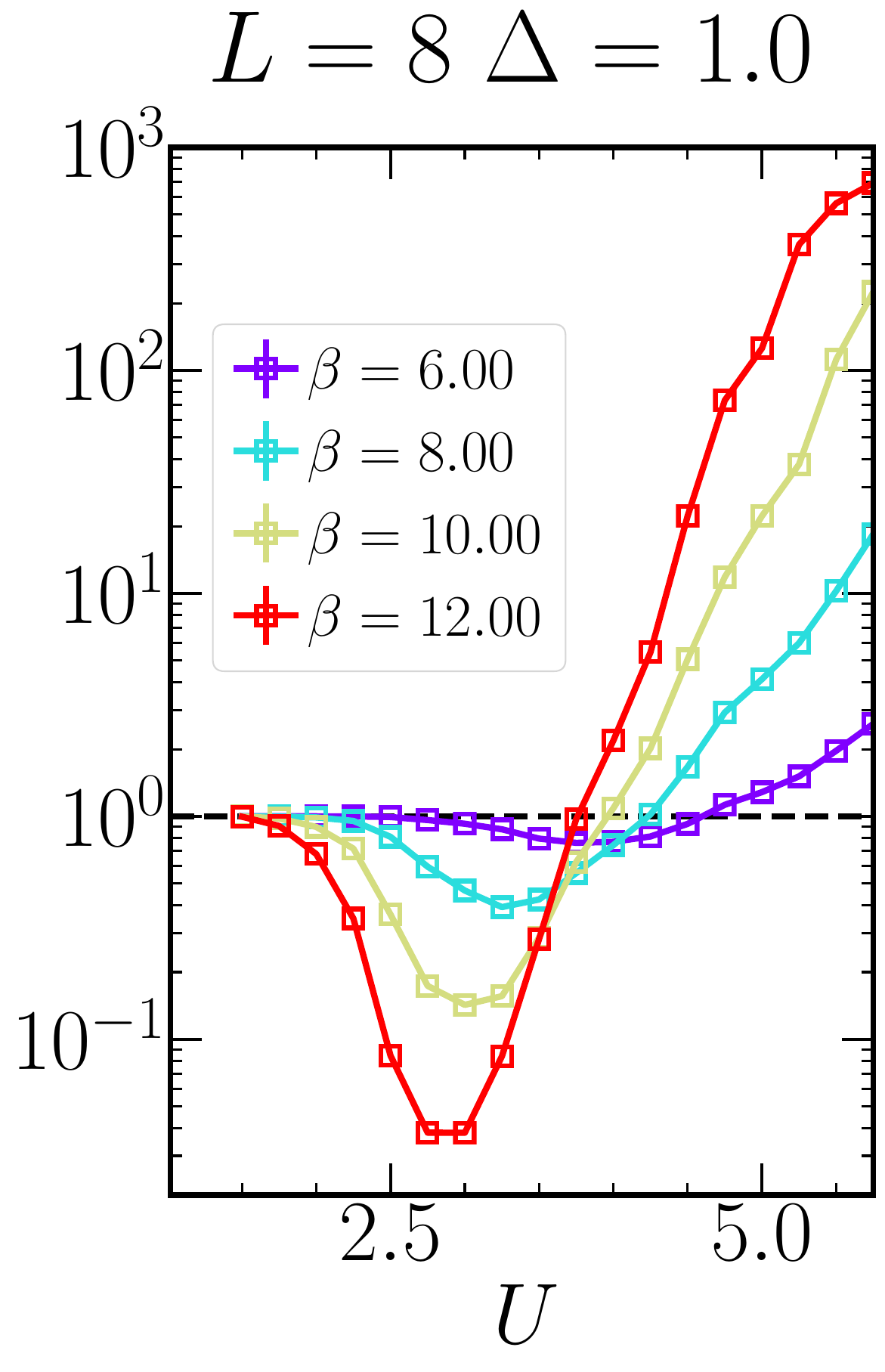}
\caption{The ratio $\langle {\cal S} \rangle (g=1) / \langle {\cal S} \rangle (g=0)$ for the square lattice Hubbard model (left), the Hubbard model on the honeycomb lattice (middle) and the ionic Hubbard model (right). The horizontal dashed black lines show $\langle {\cal S}_{1} \rangle / \langle {\cal S}_{0} \rangle = 1$, providing a reference for the other plots.
}
\label{fig:fig8}
\end{figure}

In order to summarize these detailed results, we show plots of the ratio $\langle {\cal S} \rangle (g=1) / \langle {\cal S} \rangle (g=0)$ side by side for the three models considered here in Fig.~\ref{fig:fig8}. In a sense, this ratio isolates the part of the sign due to the correct sampling (at $g=1$) from the `intrinsic' sign due to random sampling (at $g=0$). While this is a somewhat crude quantity, it provides a good first impression of the above effects and how they vary from one model to another. The left plot for the square lattice Hubbard model results clearly demonstrate how the sign gets progressively worse with increasing $g$ in the intermediate $\mu$ regime, where the ratio becomes small, and then rises up to unity again as the system approaches half filling. The slight increase above $1$ near half filling is due to the faster increase of the $g=1$ plots there, as the Mott plateau is the strongest in this case.
The middle plot shows the corresponding results for the honeycomb Hubbard model. As seen earlier in Fig.~\ref{fig:fig6}, as $g$ is lowered, the sign plots become wider and deeper systematically. This also shows up in the ratio plot, where it is almost unity for $U \lesssim 3.0$, and then increases considerably with increasing $U$. Crudely, the branching point marks the regime where random sampling starts to make the sign worse, leading up to the AFMI phase, even though this is somewhat smaller than the critical value $U_{c}$. 
In the ionic model (right column), we find that the ratio initially dips below $1$, and then rises, crossing unity approximately around $U \sim 4$ for the temperature values shown here. The initial dip is a consequence of the rightward shift in the $g=0$ plot as can be seen in Fig.~\ref{fig:fig7}, where the random sampling sign is larger than the $g=1$ value. However, as the system navigates through the CM phase and approaches the AFMI, the $g=1$ sign stabilizes again, leading to the crossing at $U<U_{c}$.  

\begin{figure}[b]
    \centering
    \includegraphics[scale = 0.27]{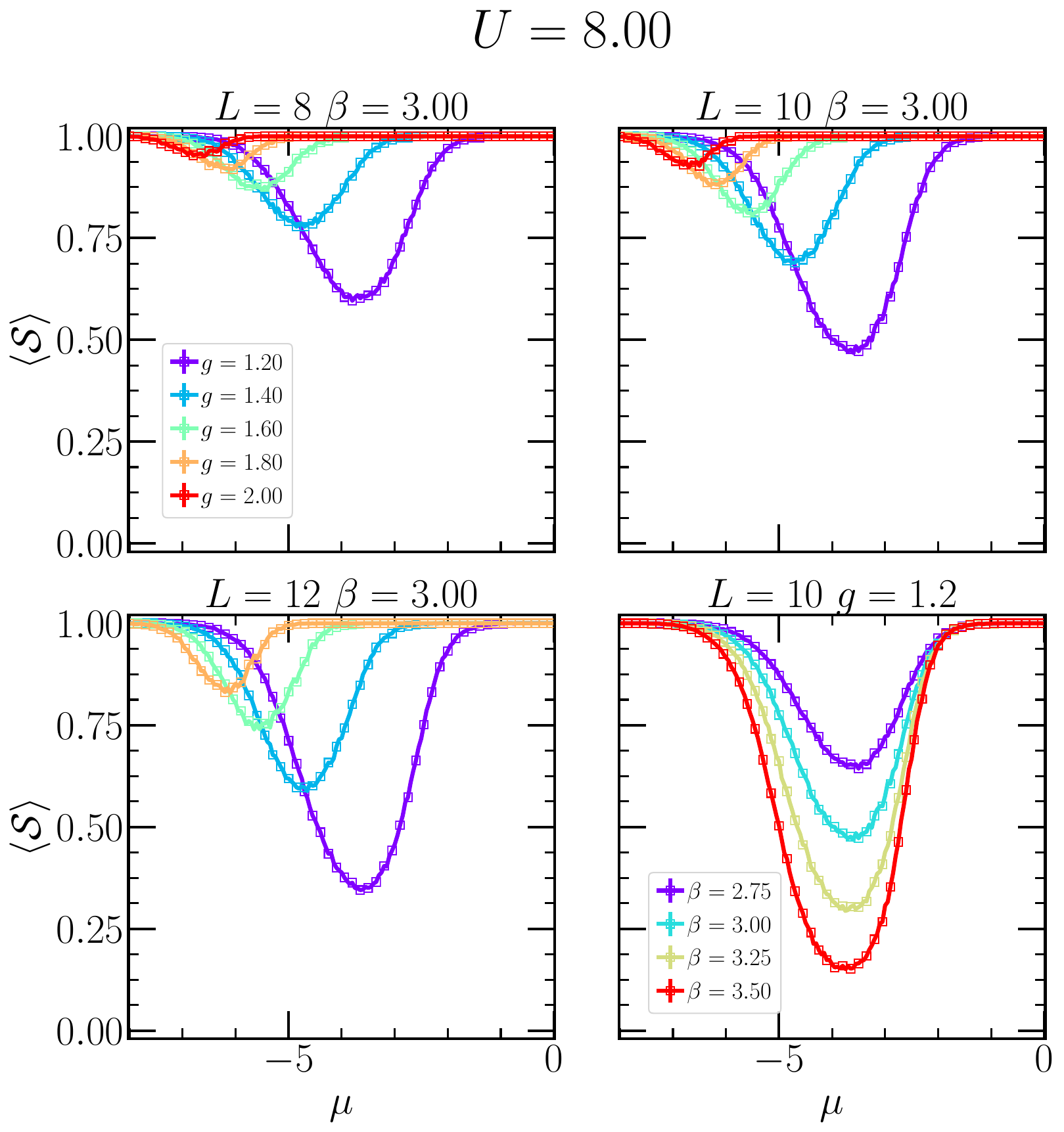}
    \caption{The average sign $\langle S \rangle$ vs.~chemical potential $\mu$ 
        at fixed $\beta = 3$ as a function of reweight factor $g>1$ at $U =8$
        on different lattice sizes, $L=8, 10$ and $12$. The SP gets worse with increasing lattice size. Increasing $g$ makes the sign shallower (i.e.~lessens the SP) and pushes the chemical potential at which the minimum 
        $\langle {\cal S} \rangle$ occurs to lower values.  With fixed lattice size $L = 10$, lowering temperature results in smaller
        $\langle {\cal S} \rangle$.
 }
    \label{fig:fig9}
\end{figure}



\section{Oversampling}  \label{sec:oversampling}

In the preceding sections we have considered $0<g<1$, which interpolates from
random sampling of the HS fields ($g=0$) and the exact simulation at $g=1$.
As we have discussed, this allows us to separate two `sources' of the SP:
the inherent possibility of negative determinants for {\it any}, i.e.~randomly selected,
fermionic matrices, and the preferential likelihood of such matrices induced
by the sampling.  Here we explore an additional issue, namely the behavior for $g>1$.
We are indirectly motivated by the `successive over-relaxation method'~\cite{young54,hackbusch94} to solve linear equations
which suggests an iterative move from $k$ to $k+1$ {\it beyond} the value at 
$k+1$ initially computed.  The analogy here is that $g>1$ samples the HS field
beyond what the $g=1$ determinant suggests, but, admittedly, we are also
motivated by plain old curiosity.
We focus exclusively on the square lattice.

In Fig.~\ref{fig:fig9}, we show plots of $\langle {\cal S} \rangle$ vs.~$\mu$ with $1<g<2$ for three different lattice sizes in three panels as marked. We find that the minimum in the sign becomes shallower and drifts rapidly to the left (for $\mu<0$, the plots are symmetric about half filling) as $g$ is increased. The observations in the previous sections immediately suggest an explanation: increasing $g$ pushes the system towards stronger coupling, resulting in a larger Mott plateau which improves the sign and pushes the onset of the SP to more negative $\mu$, beyond the Mott plateau. As expected, the SP becomes progressively worse with increasing lattice size, as the panels clearly demonstrate. The bottom right panel explores the temperature dependence of the sign at $g = 1.2$ on a $10 \times 10$ lattice. We find a broad minimum in $\langle {\cal S} \rangle$ which grows deeper with reducing temperature, as expected. As $g$ is increased, the Mott plateau grows progressively bigger and by $g=2.0$ (not shown), the sign is practically saturated at unity throughout
the range shown, $-8 < \mu < 0$. 

\begin{figure}[t]
    \centering
    \includegraphics[scale = 0.27]{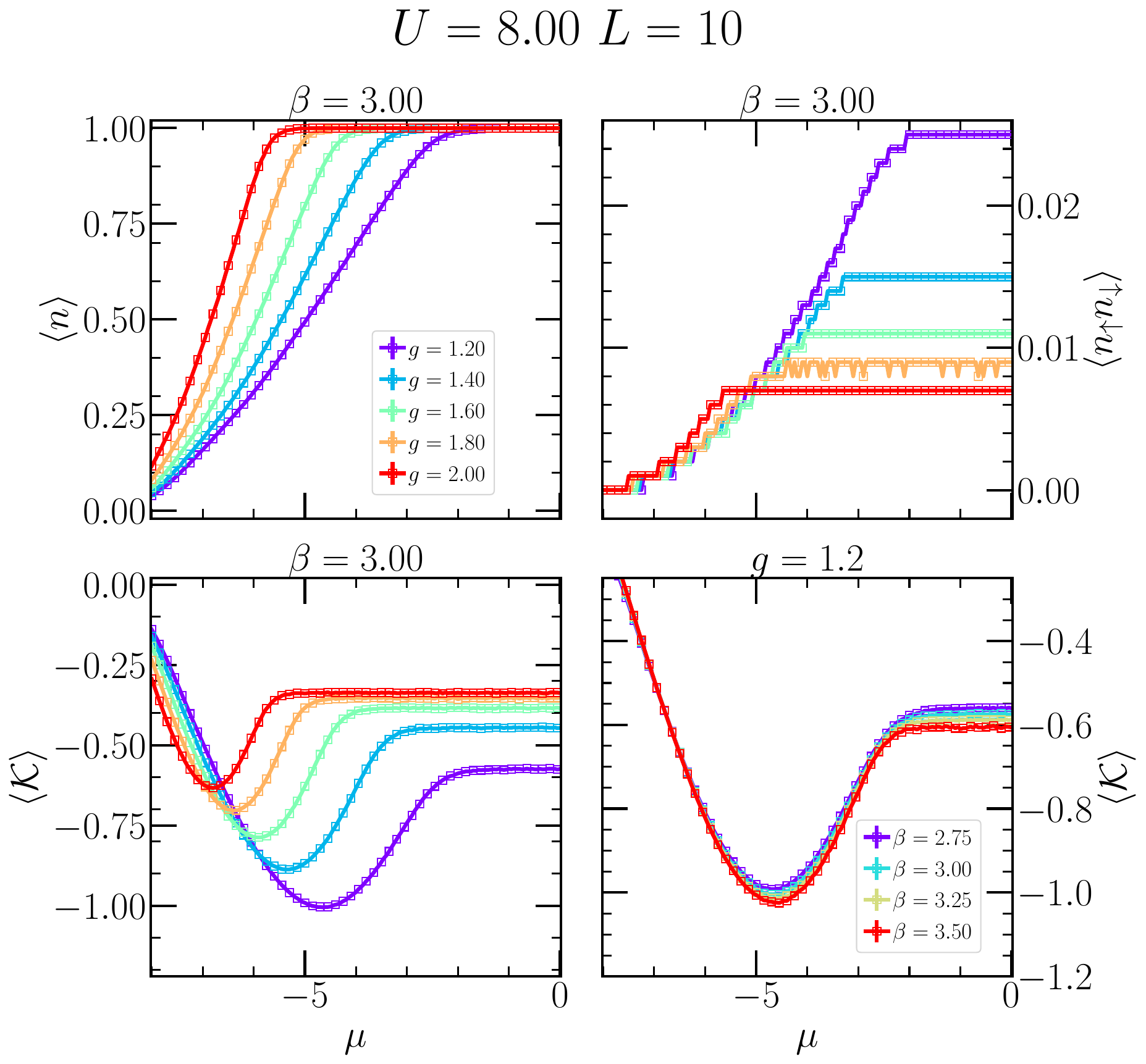}
    \caption{Oversampling case: electron density $\langle n \rangle$, double occupancy $\langle n_{\uparrow} n_{\downarrow} \rangle$ and kinetic energy $\langle {\cal K} \rangle$ vs. $\mu$ as a function of $g$ at $U =8$ on a $10 \times 10$ square lattice. Bottom right panel displays the temperature dependence of $\langle {\cal K} \rangle$ at $g$ = $1.2$ (density and double occupancy have negligible lattice and temperature dependence in this regime).}
    \label{fig:fig10}
\end{figure}

In Fig.~\ref{fig:fig10}, we show the behaviour of local physical variables such as the number density $\langle n \rangle$, the double occupancy $\langle n_{\uparrow} n _{\downarrow} \rangle$ and the kinetic energy $\langle {\cal K} \rangle$ vs. $\mu$ in this $g>1$  regime. As the Mott insulator becomes more extensive with increasing $g$, we find that $\langle n \rangle$ saturates at progressively more negative values of $\mu$. The double occupancy and the kinetic energy both reduce in magnitude as the insulating system increasingly favors single occupancy. The bottom right panel plots $\langle {\cal K} \rangle$ for varying temperatures at $g = 1.2$, demonstrating a rather weak dependence on temperature. Similarly, we find negligible dependence on temperature and lattice size for all these variables in this regime (data not shown here). It is quite striking that while the sign itself varies considerably with temperature, the physical variables show little change in comparison; a peek at Fig.~\ref{fig:fig3} reveals that the same observation is largely true in the `undersampling' case ($g<1$) as well.

\begin{figure}[t]
    \centering
    \includegraphics[scale=0.29]{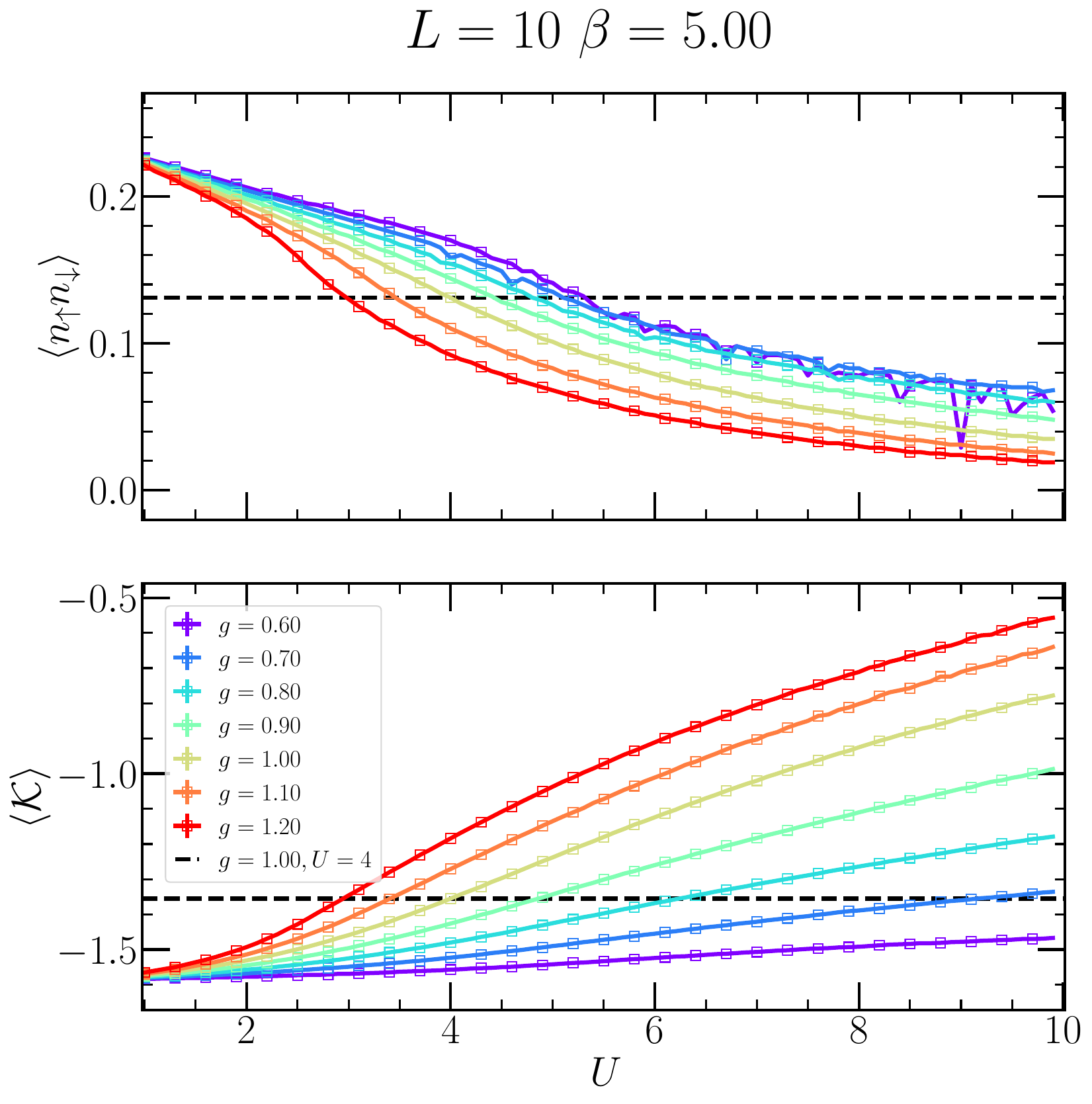}
    \caption{Double occupancy $\langle n_{\uparrow}n_{\downarrow} \rangle$  (top) and kinetic energy $\langle \mathcal{K} \rangle$ (bottom) vs. on-site interaction strength $U$ using different reweight factors $g$.  
    Here $L = 10 $ and $\beta =5$. The dashed horizontal line intersects the $g=1$ curve at $U=4$. To get the same value of 
$\langle n_{\uparrow}n_{\downarrow} \rangle$ for $g=0.7$ requires $U = 5.1$, and for $g=1.2$ requires $U = 3$. This implies $U_{\rm eff}=4$ for $g=1.2, U = 3.0$ and for $g=0.7, U = 5.1$.
A similar analysis of $\langle \mathcal{K} \rangle$ implies that
$U_{\rm eff}=4$ for $g=1.2$, $U = 2.8$ and for $g=0.7$, $U = 9.2$.}
    \label{fig:fig11}
\end{figure}

\section{Renormalized Coupling}  \label{sec:renorm}


\begin{figure}[t]
    \centering
    \includegraphics[scale = 0.29]{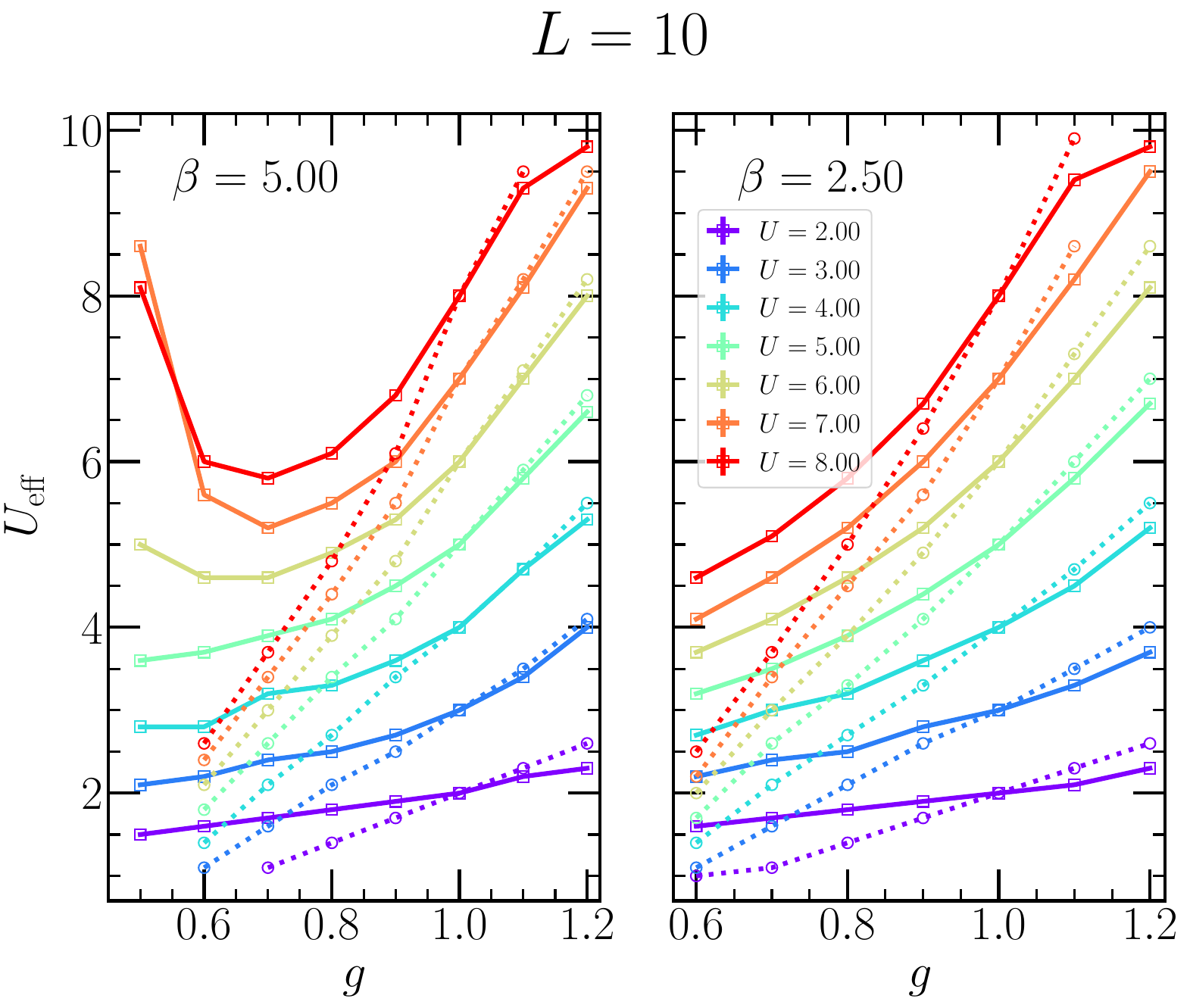}
    \caption{Effective on-site interaction $U_{\rm eff}$ determined from double occupancy $\langle n_{\uparrow} n_{\downarrow} \rangle$ (solid lines)  and kinetic energy $\mathcal{K}$ (dotted lines) vs.~reweight factor $g$. Inverse temperature is fixed at $\beta = 5$ (left panel) and $\beta = 2.5$ (right panel), and data was extracted using an $L=10$ lattice. For $g>1$ the inferred values are roughly equal, while for small $g$
    there is a marked disagreement.  $U_{\rm eff}$ seems to be roughly temperature independent except for small $g$ and large $U$.}
    \label{fig:fig12}
\end{figure}

It is useful to quantify these statements concerning the effect
of $g$ on the physics.  To do this, we
determine an effective (renormalized)
interaction strength $U_{\rm eff}$ due to $g \neq 1$.  We proceed as follows: 
for a given $g$ and $U$ we find the value of the repulsion, $U_{\rm eff}$, 
which at $g = 1$ yields the same value for local observables including 
the double occupancy $\mathcal{D} = \langle n_{\uparrow} n_{\downarrow} \rangle$ or 
the kinetic energy $\langle \mathcal{K} \rangle$. In Fig.~\ref{fig:fig11}, 
we show 
$\mathcal{D}$ and $\mathcal{K}$ 
as functions of $U$
at half-filling ($\langle n \rangle=1$) on a $N = 10 \times 10$ lattice.  
For a given value of $U$ at $g=1$, a horizontal cut gives the value of $U_{\rm eff}$ from the intersections with the curves at
$g \neq 1$.  Comparison of 
$\mathcal{D}$ and $\mathcal{K}$ 
reveals that the results are not universal- the inferred $U_{\rm eff}$ depends on the observable especially for $g<1$.

In Fig.~\ref{fig:fig12}, we compare the values of $U_{\rm eff}(g)$ determined in the manner described from  Fig.~\ref{fig:fig11}. 
In general, the double occupancy $\mathcal{D}$ tends to predict larger $U_{\rm eff}$ than the kinetic energy
$\mathcal{K}$ when $g<1$, but the values are similar for $g>1$.

\section{Conclusions}  \label{sec:conclusions}

In this paper, we have argued that a simple view of the SP as originating in the independent winding of fermionic world lines across the space-time lattice
is incomplete.  Instead, by studying the Hubbard model in three different contexts, namely, on a square lattice, on a honeycomb lattice, and with an external staggered potential, the SP has been shown to depend non-trivially on the manner in which the fermion determinants select the Hubbard-Stratonovich fields. While the square lattice results show that the SP predominantly originates from determinants steering the simulation to low sign regions of phase space, the honeycomb and ionic results demonstrate that random sampling makes the sign worse, especially leading up to the AFMI phase. Given the few models considered here, it is difficult to infer a general rule that determines the extent of the different contributions to the SP under a given set of circumstances. However, as seen above, when the determinants guide the system towards a protected particle-hole symmetric point, such as the AFMI state, the SP is typically reduced. Extensions of this work to other models should be able to shed more light on this issue.

As we have seen, shifting $g$ away from $g=1$ explicitly changes the values of physical observables. Indeed, it has the effect of pushing the system to weaker coupling in the sense that the effects of correlation such as the Mott gap and the suppression of double occupancy and hopping
are reduced. Similarly, in the `oversampling' case, with $g>1$, the effective coupling is larger. The resulting reinforcement of  the Mott plateau in the square lattice case rapidly mitigates the SP. An explicit  calculation of the effective coupling puts these observations on a more quantitative footing. However, unlike an analogous analysis for the renormalization of the electron-phonon coupling in a model which includes on-site interactions, where the effective coupling was the same for different observables~\cite{feng21}, here only whether $U_{\rm eff}$ is reduced or enhanced relative to $U$ is universal.  The quantitive value of $U_{\rm eff}$ can vary markedly depending on which observable is analyzed.

Our procedure allows us to de-convolve the ways in which the sign problem arises,
and the alteration of the underlying physics provides us with qualitative insight into
the observed trends.  It is, of course, possible to formulate a protocol in which the alteration
of the weight ${\cal W}$ by $g \neq 1$ is compensated by including an appropriate  factor ${\cal W}^{1-g}$
in the measurement of physical observables.  In this case, expectation values would be unchanged,
but the error bars would be altered.  Intuitively, setting $g \neq 1$ seems likely to lead to 
{\it less efficient} simulations as it violates the spirit of proper importance sampling.

A final comment concerns connections to what we have explored here to the
`rational hybrid Monte Carlo' (RHMC) algorithm~\cite{clark07} widely used in lattice gauge theory.
In RHMC the fermion determinant is split into many pieces,
${\rm det} M = \big(\, {\rm det}( M^{1/n}) \,\big)^n$.
The purpose there has nothing to do with sign problem, but rather to make the 
computation of a `pseudofermion' approximant to the determinant more well conditioned,
allowing larger step sizes in the integration of the equations of motion.
Nevertheless, it is interesting that the $1/n$ factor plays a very similar role
to the $g$ considered here.  The difference of course is that our approach
no longer simulates the original model when $g<1$ because we do not
introduce an increased number $n=1/g$ of copies of the determinant.

\section{Acknowledgments}
\noindent
R.T.S.~was supported by the grant DE‐SC0014671 funded by the U.S.~Department of Energy, Office of Science.
R.M.~acknowledges support from the National Natural Science Foundation of China (NSFC) Grants No.~U1930402, No.~11974039, 
No.~12050410263. 
B.X.~was supported by the Flatiron Institute.  The Flatiron Institute is a division of the Simons Foundation.
Computations were performed on the Tianhe-2JK at the Beijing  Computational  Science  Research  Center.
 










\appendix


\section{Sign of Individual Spin Species}
\label{sec:app1}

\begin{figure}[t]
\includegraphics[scale=0.27]{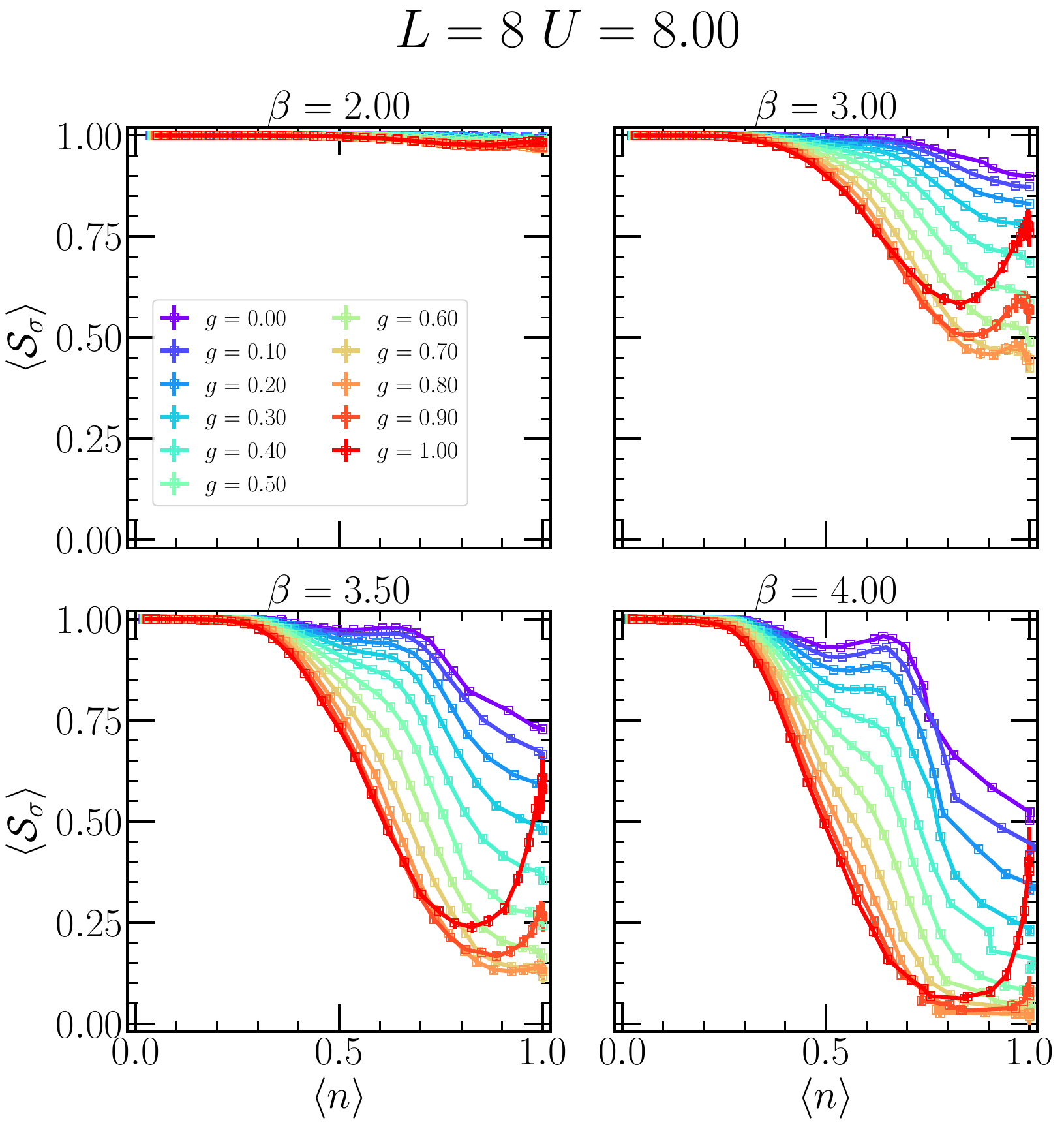}
\caption{The average sign of the matrices for individual spins, $\langle {\cal S}_{\sigma} \rangle$, vs. $\langle n \rangle$ at $U=8$ as a function of $g$ on an $8\times8$ lattice for four values of $\beta$. As seen for the total sign in the main text, the minimum in $\langle {\cal S}_{\sigma} \rangle$ becomes wider and deeper with increasing $\beta$. The non-monotonic behaviour of the sign with reducing $g$, where ${\cal S}_{\sigma}$ becomes worse initially, and then gets progressively better, is also evident here. In contrast to the total sign, however, the spin resolved sign is not constrained to be equal to $1$ at half filling, and thus the sharp upturn seen in Fig.~\ref{fig:fig1} near $\mu = 0$ is absent here for most $g$ values.
}
\label{fig:figs1}
\end{figure}

Fig. ~\ref{fig:figs1} shows the behavior of the sign of the individual weight matrices, $\langle {\cal S}_{\sigma} \rangle$ with the rescaling parameter $g$ at four values of the temperature. We find that the spin resolved sign shows very similar qualitative behaviour to the total one except near half filling, where the constraint of particle-hole symmetry that led to the sign approaching unity near half filling is no longer present for the spin resolved quantity. Interestingly, the plots with $g \gtrsim 0.8$ still show an upturn near half filling, even though they do not reach the maximum value. The non-monotonicity of the sign noted in Fig.~\ref{fig:fig1} with reducing $g$, where it becomes worse initially before increasing again, is also visible here for $\langle {\cal S}_{\sigma} \rangle$.



\section{Number density at different lattice sizes}
\label{sec:app2}

\begin{figure}[t]
\includegraphics[scale=0.27]{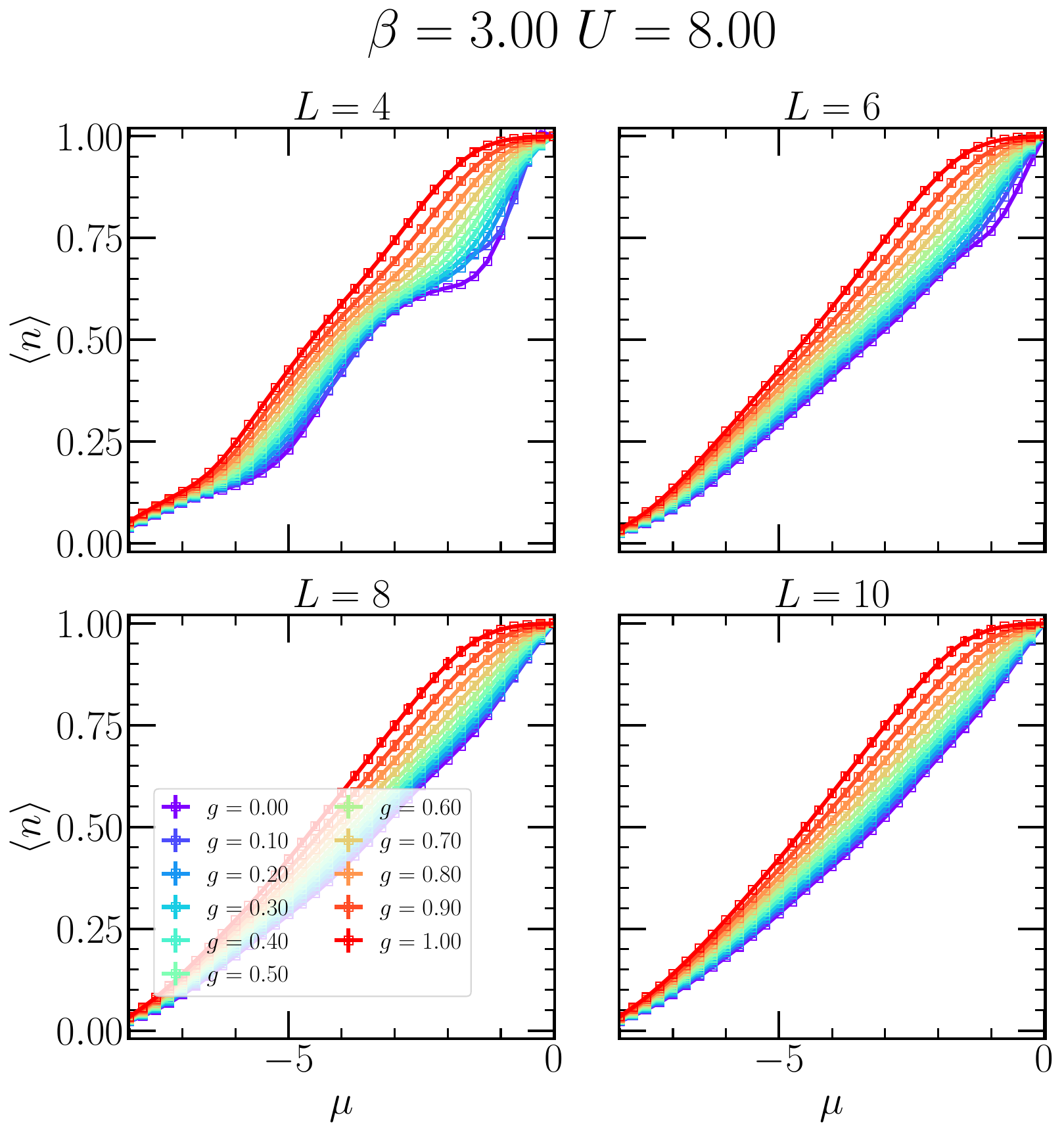}
\caption{Density $ \langle n \rangle$, vs. $\mu$ at $U=8$, $\beta = 3.0$ on an $8 \times 8$ lattice as a function of $g$ for four different values of the lattice size $L$. The oscillatory behaviour at low $g$ in the $4 \times 4$ lattice is due to  the re-emergence of the shell effect in small lattices. As explained in detail in the main text, a completely random sampling of the Hubbard Stratonovich fields in QMC is more involved than a standard rescaling of $T$; in this case, pushing the system to weaker coupling by reducing the effect of $U$, resulting in a reappearance of finite size effects originally suppressed by the interaction. 
}
\label{fig:figs2}
\end{figure}

In Fig.~\ref{fig:figs2}, we show the number density vs. chemical potential for different lattice sizes $L$. Apart from the usual features of the Mott saturation near half filling at high values of $g$, and its absence at low values of the rescaling parameter that we have already noted in the main text, we find that the finite size effects, evidenced in the oscillations of $\langle n \rangle$ for $g \sim 0$, are considerably enhanced at lower lattice sizes, as expected.


\section{Sign vs. lattice size}
\label{sec:app3}


\begin{figure}[h]
\includegraphics[scale=0.27]{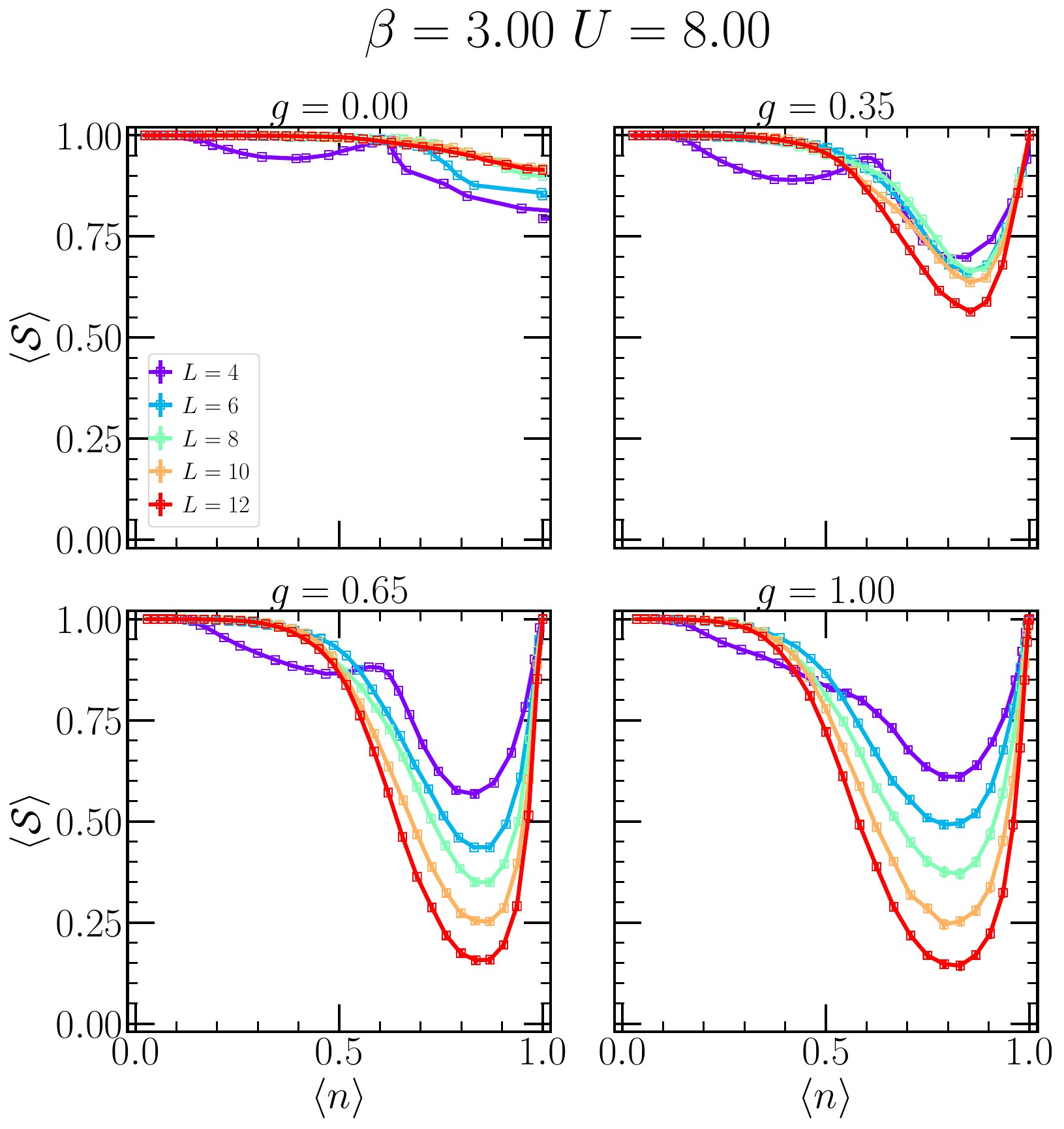}
\caption{$\langle {\cal S} \rangle$ vs. $\langle n \rangle$ at $\beta=3$, $U=8$ as a function of $L$ for different $g$ values. Main observation is the finite size oscillations for $L=4$, with the sign reaching a maximum at the `magic density' $\langle n \rangle \sim 0.625$ associated to a closed shell filling for this system size. 
}
\label{fig:figs3}
\end{figure}

\begin{figure}[b]
\includegraphics[scale=0.27]{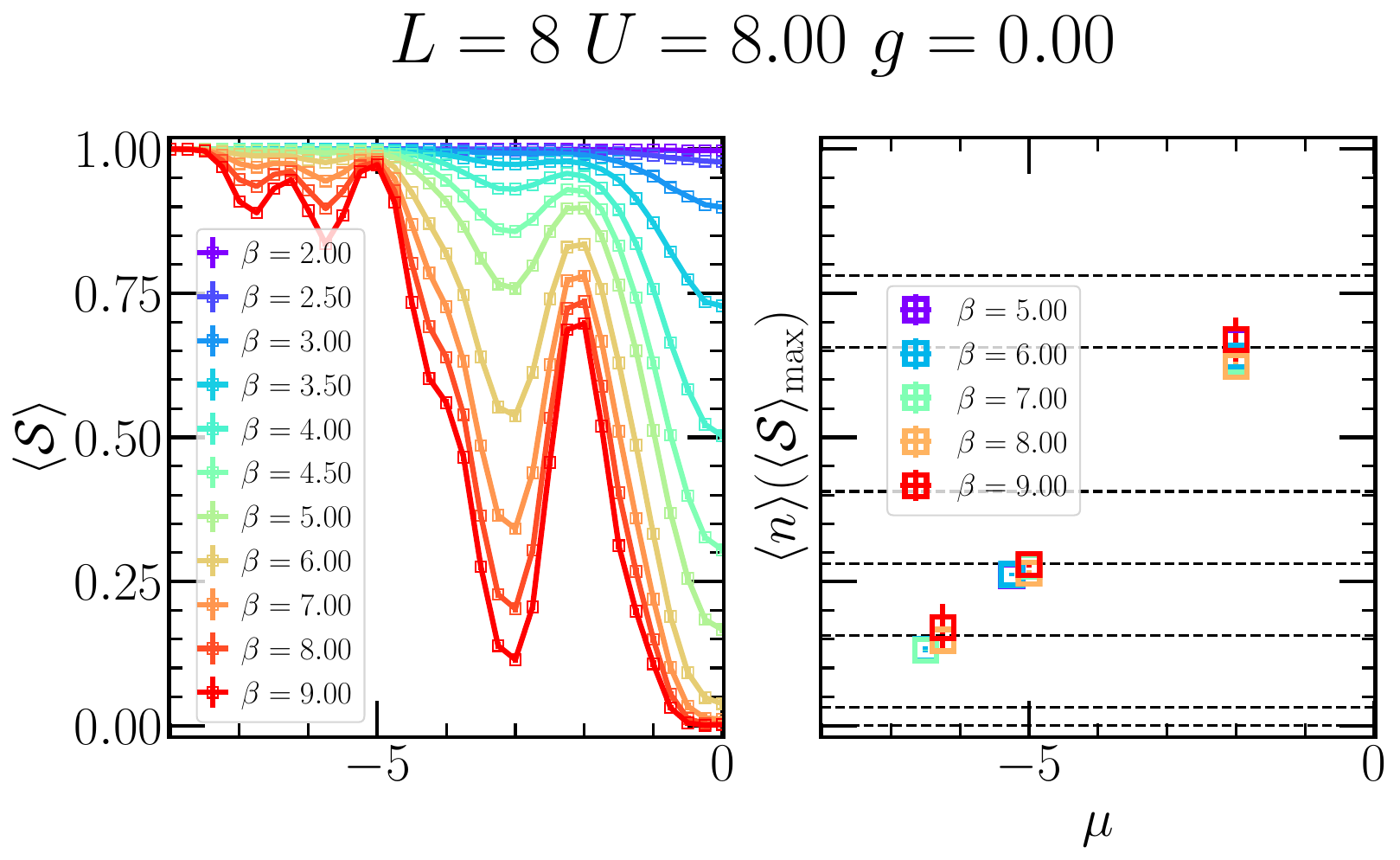}
\caption{Left: $\langle {\cal S} \rangle$ vs. $\mu$ at $g=0$, $U=8$, $L=8$ for
different $\beta$ values. As $\beta$ is increased above $4$, $\langle {\cal S} \rangle$
 shows strong oscillatory behaviour reminiscent of shell effects. Right: values of $\langle n \rangle$ at the maxima
 of the sign. Dotted black lines mark the `magic values' of the density at this lattice
 size (see text). 
 }
\label{fig:figs4}
\end{figure}

\begin{figure}[h]
\includegraphics[scale=0.27]{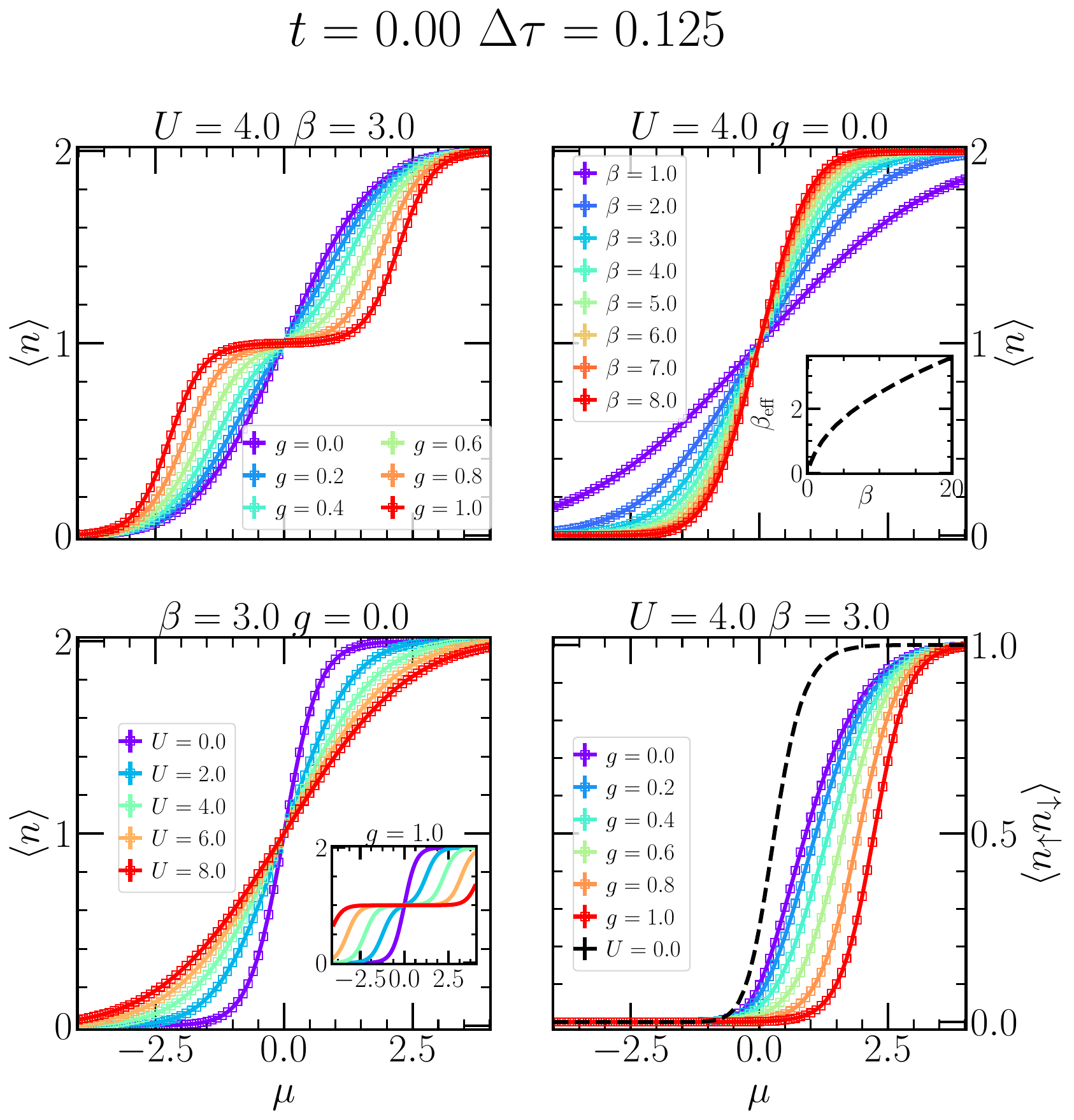}
\caption{The $t=0$ limit. Top left panel shows the number density $\langle n \rangle$ vs. chemical potential $\mu$ at different $g$ values, showing the `Mott plateau' near $g=1$ disappearing as $g \rightarrow 0$. Top right panel plots $\langle n \rangle$ vs $\mu$ for $U=4.0$, $g=0$ at different $\beta$. Inset fits these curves to a Fermi function with $\beta_{{\rm eff}}$. Bottom left shows the same for fixed $\beta = 3.0$, $g=0$ at different coupling $U$, with inset plotting the same at $g=1$. Final plot shows the double occupancy, $\langle n_{\uparrow} n_{\downarrow} \rangle $, for different $g$ values, along with the $U=0$ plot for comparison.}
\label{fig:figs5}
\end{figure}

In Fig.~\ref{fig:figs3}, we show the sign $\langle {\cal S} \rangle$ plotted against the density $\langle n \rangle$ as a function of the lattice size $L$ for four different values of $g$. Aside from the usual observations, we clearly see the finite size oscillations for the $L=4$ plots, with the sign reaching a maximum at the `magic density' $\langle n \rangle \sim 0.625$, corresponding to a closed shell filling for this system size. 


\section{Finite size oscillations in the $g=0$ limit}
\label{sec:app4}



In Fig.~\ref{fig:figs4}, we reinforce the statements made in several parts of the main text that as we approach $g=0$, the effect of $U$ is reduced on certain physical attributes of the system, and noninteracting features, such as finite size shell effects, usually washed out by the strong interaction, are unmasked. The left panel shows the sign at $g=0$ plotted for a large number of temperature values going down to $\beta=9$. Strong oscillations in $\langle {\cal S} \rangle$ are observed as lower temperatures are reached. 
The peak positions in this limit are connected to the closed-shell densities for this system size, as confirmed by the right panel, which plots the positions of the densities corresponding to the maxima of the sign at low temperatures. This is more evident when comparing it to the dotted horizontal lines, which mark the values of the closed shell densities for $8\times8$ lattices. 


\section{The $t=0$ limit}
\label{sec:app5}


Here, we discuss the strong coupling ($t=0$) limit in detail. As mentioned briefly in the main text, this limit can be solved analytically at $g=1$ ($g=0$ is trivial, of course), whereas for a general value of $g$, the solution cannot be written down in a closed form. 

The partition function for a general value of $g$ is given by

\begin{align}
 {\cal Z} &= {\cal C}^{N L_{\tau}}~\prod_{i} \sum_{ \{ h_{i} \}}  \Big\{ \prod_{\sigma} \Big( 1 + e^{\beta \mu + \alpha \sigma \sum_{l} h_{i}(l)} \Big) \Big\}^{g} \end{align}

For $g = 1$, this is trivially easy to calculate

\begin{align}
{\cal Z} &= {\cal C}^{N L_{\tau}} \prod_{i} \sum_{ \{ h_{i} \}} \Big\{ e^{\beta \mu} \Big(e^{\alpha \sum_{l} h_{i}(l)} + e^{-\alpha \sum_{l} h_{i}(l)} \Big) + \nonumber \\ &~~~~~~~~~~~~~~~~~~~~~~~~~~~~~~~~~~~~~~~~~~~~~~~~~~~~~~1+ e^{2\beta \mu} \Big\}  \nonumber \\
&={\cal C}^{N L_{\tau}} \prod_{i}~ \Big\{ 2 e^{\beta \mu} \Big(2 {\rm cosh}(\alpha) \Big)^{L_{\tau}} + 2^{L_{\tau}} \Big(1 + e^{2 \beta \mu} \Big) \Big\} \nonumber \\
&= \prod_{i} e^{-\beta U/4}~\Big( 1 + e^{2 \beta \mu} + 2 e^{\beta (\mu + U/2)} \Big)
\end{align}

On the other hand, the general case with $g \neq 1$ cannot be solved to yield an analytical expression as above, and have to be computed numerically. However, the calculations may be simplified by noting that the summand is a function of $\sum_{l} h_{i}(l)$ for any given site $i$. Since each site is independent, we will just choose one site and drop the label $i$ in what follows below.
Each of the HS fields $h(l)$ takes values $\pm 1$. Hence, for a general situation where $n$ of them are $-1$ and the rest $+1$, the sum $\sum_{l} h(l) = L_{\tau} - 2n$. The degeneracy of such a situation is given by $\Mycomb[L_{\tau}]{n} = \frac{L_{\tau} !}{n !~ (L_{\tau} - n) !}$, the number of ways of choosing $n$ variables out of $L_{\tau}$. Thus, the single site partition function, ${\cal Z}_{s}$, may be rewritten as

\begin{align}
  {\cal Z}_{s} = {\cal C}^{L_{\tau}} \sum^{L_{\tau}}_{n=0} 
  \Mycomb[L_{\tau}]{n} \Big\{ \prod_{\sigma} \Big( 1 + e^{\beta \mu + \alpha \sigma (L_{\tau} - 2n)} \Big) \Big\}^{g}
\end{align}

The equal time Green's function ${\cal G}_{\sigma}(\tau, \tau)$ is independent of the imaginary time co-ordinate $\tau$ and is given by ${\cal G}_{\sigma}(\tau, \tau) = 1 / (1 + e^{\beta \mu + \alpha \sigma \sum_{l} h(l)})$. For any given configuration $\{ h\}$, this is simply the equivalent of the non-interacting expression $\big(1 - 1 / (1 + e^{\beta \epsilon_{\sigma}}) \big)$, with $\epsilon_{\sigma} = - \alpha \sigma \sum_{l} h(l) - \mu$.
Calculations of expectation values of variables ${\cal A}({\cal G})$, expressed in terms of the equal time Green's function, may be performed in a similar manner to ${\cal Z}_{s}$:

\begin{align}
    \langle {\cal A} \rangle = {\cal Z}^{-1} ~{\cal C}^{L_{\tau}} \sum^{L_{\tau}}_{n=0} & 
  \Mycomb[L_{\tau}]{n} ~{\cal A}({\cal G}) \nonumber \\ &\Big\{ \prod_{\sigma} \Big( 1 + e^{\beta \mu + \alpha \sigma (L_{\tau} - 2n)} \Big) \Big\}^{g}
\end{align}

In Fig.~\ref{fig:figs5}, we show the number density $\langle n \rangle$ and the double occupation $\langle n_{\uparrow} n_{\downarrow} \rangle$ as a function of $\mu$ for various values of $g$, $U$ and $\beta$. As seen in the general cases with finite coupling in the main text, we find that the `Mott' insulator at $g=1$ gradually disappears as $g$ is dialled down. Similarly, the double occupancy in the bottom right panel also shows that the effect of $U$ becomes weaker as $g$ is reduced. However, as the $U=0$ black line indicates, even $g=0$ is very different from the non interacting case. Top right and bottom left panels show the $g=0$ plots for $\langle n \rangle$ for different values of $\beta$ (fixed $U$) and different $U$ (fixed $\beta$) respectively. While these can be nominally fitted to noninteracting Fermi functions, the effective $\beta$ (top right inset) is finite (the classical $\beta \rightarrow g \beta$ scaling implies infinite temperature) but is smaller than the actual value. Already at the single site level, this demonstrates the non-trivial effect that tuning $g$ has on the physics of the interacting system.

\bibliography{inferringphysicssign}

\end{document}